\documentclass[a4paper,11pt]{elsarticle}

\newcommand{\com}[1]{}

\usepackage{graphicx}
\usepackage{amssymb}
\usepackage{amsmath}
\usepackage{amsfonts}
\usepackage{geometry}
\usepackage{here}
\usepackage[]{hyperref}
     \usepackage{epsfig}
\usepackage{array}
\usepackage{color}
      \usepackage{bm}
\newcolumntype{C}[1]{>{\vspace{1.5pt}\centering\let\newline\\\arraybackslash\hspace{0pt}}m{#1}}

\newif\iffigsdirectory

\begin{document}

\title{The Schr\"odinger-Langevin equation with and without thermal fluctuations.}

\author{R. Katz}
\ead{roland.katz@subatech.in2p3.fr}

\author{P. B. Gossiaux}
\ead{Pol-Bernard.Gossiaux@subatech.in2p3.fr}

\address{SUBATECH (UMR 6457)\\ 
Ecole des Mines de Nantes, CNRS/IN2P3, Universit\'e de Nantes 
\\ 4 rue Alfred Kastler, 44307 Nantes cedex 3, France}



\begin{abstract}
The Schr\"odinger--Langevin equation (SLE) is considered as an effective open quantum system formalism suitable for phenomenological applications involving a quantum subsystem interacting with a thermal bath. We focus on two open issues relative to its solutions: the stationarity of the excited states of the non-interacting subsystem when one considers the dissipation only and the thermal relaxation toward asymptotic distributions with the additional stochastic term. We first show that a proper application of the Madelung/polar transformation of the wave function leads to a non zero damping of the excited states of the quantum subsystem. We then study analytically and numerically the SLE ability to bring a quantum subsystem to the thermal equilibrium of statistical mechanics. To do so, concepts about statistical mixed states and quantum noises are discussed and a detailed analysis is carried with two kinds of noise and potential. We show that within our assumptions the use of the SLE as an effective open quantum system formalism is possible and discuss some of its limitations.%
\end{abstract}


\maketitle

\section{Introduction}\label{Sec1Intro}

In classical mechanics, the influence of a thermal environment (bath) on a Brownian particle (subsystem) is well described by the Langevin dynamics within the Newtonian framework. The subsystem thermalization is obtained from the balance of two forces (friction and stochastic) which generate irreversible energy exchanges between the two systems.
To search for the corresponding description in quantum mechanics is a crucial issue both for the understanding of quantum fundamentals and in many branches of applied physics (where the quantum systems can never be isolated), such as in quantum diffusion and transport \cite{Bhattacharya:2011,PhysRevLett.105.136101,Sanz:2013}, quantum optics \cite{Degman:1998,Horowitz:2012}, heavy ion scattering \cite{Gross:1978,Hamdouni:2010,Akamatsu:2011se}, quantum computers and devices \cite{Nielsen:2000,Pekola:2015,Henriet:2014}. Unfortunately, the Langevin dynamics -- or more generally energy dissipation -- cannot be introduced easily in the common quantum formalism, as no direct canonical quantization of a Hamiltonian can describe irreversible phenomena \cite{Lindblad:1976wv}. \\

In the present work, we focus on a possible Langevin-like extension of the fundamental Schr\"odinger equation, the so-called Schr\"odinger--Langevin equation (``SLE")
\begin{eqnarray}\label{SLeq}
i\hbar\frac{\partial \psi(x,t)}{\partial t}&=& \Bigg[ H_0 + \hbar A\Big(S(x,t)-\int\psi^*S(x,t)\,\psi \,\,dx\Big) -x F_R(t)\Bigg]\,\psi\,,
\end{eqnarray}
where $A$ is the friction coefficient, $S$ the (real) phase of the wave function, $F_R(t)$ a fluctuation operator and $H_0$ the usual isolated-subsystem Hamiltonian,
\begin{eqnarray}\label{NoThermHamil}
H_0=-(\hbar^2/2m)\nabla^2+V_{\rm ext}({x}).
\end{eqnarray}
The SLE was first proposed by Kostin \cite{Kostin:1972} from an identification with the Langevin equation for Heisenberg operators, the so-called Heisenberg--Langevin equation (``HLE"),
\begin{equation}\label{HLeq}
\dot{P}=F_{\rm ext}(X)-A\,P+F_R(t)\qquad\mbox{and}\qquad \dot{X}=P/m \,.
\end{equation} 
The latter is derived within the common subsystem plus bath approach\footnote{In the common approach, the subsystem plus bath is considered as a whole conservative system. By integrating out the bath degrees of freedom, one obtains the dissipative evolution of the subsystem \cite{Weiss:2012}.} from a simple model of the bath \cite{Senitzky,Ford:1965,Caldeira:1983} -- a thermal ensemble of oscillators linearly coupled to the subsystem -- and has proven to be a suitable framework to study Brownian motion. The practical application of the HLE is nevertheless limited by its non-commutating operator nature. The SLE has also been derived within many other frameworks and non-standard quantization procedures to describe either pure dissipation \cite{Kan:1974,Razavy:1977,Nassar:1985a,Doebner:1997ry,Fulop:1998rk,Wysocki:2000,Garas:2013} or the case of a Brownian motion \cite{Yasue:1978bx,Ruggiero:1985,Bolivar:1998}. The SLE includes a thermal fluctuation term $-x F_R(t)$ and a dissipative term under its hydrodynamic formulation \cite{Nassar:1985a,Garas:2013,Yasue:1978bx}
\begin{eqnarray}\label{DissOper}
\hbar A(S(x,t)-\left<S\right>),
\end{eqnarray}
where the phase $S$ is chosen according to a prescription which will be discussed in Sec.~\ref{PolarPres}. Even though the dissipative term is nonlinearly (logarithmically) dependent on the wave function, it still corresponds to a linear ohmic friction (i.e.~proportional to the particle velocity). A nonlinear friction can be obtained by extending Kostin derivation to a nonlinear coupling \cite{Bargueno:2014,Vargas:2015}. The SLE exhibits interesting properties: unitarity is preserved at all times \cite{Garas:2013}, the uncertainty principle is always satisfied\footnote{As opposed to other models like the Caldirola--Kanai equation \cite{CaldiKan} without fluctuations, the Wigner--Moyal equation with classical Fokker--Planck terms or the quasiclassical HLE \cite{Weiss:2012}.} \cite{Dekker:1981,Haas:2013,Sanin:2014} and the superposition principle is violated due to the nonlinearities (which might not be a problem per se for dissipative equations \cite{Leggett:1980,Caldeira:1985}). 
Thanks to its straightforward formulation -- in principle only two ``classical" parameters need to be known: the friction coefficient $A$ and the bath temperature $T_{\rm bath}$ -- and its numerical simplicity, the SLE can be considered as a solid candidate for effective description of open quantum systems hardly accessible to quantum master equations \cite{Weiss:2012,LindbladG:1976}. Indeed, in a number of complex applications, defining the bath/interaction Hamiltonian and calculating the Lindblad operators without too many approximations is rather complicated, and some effective approaches -- possibly of the Langevin type -- are unavoidable \cite{Weiss:2012,Breuer:2002}. Because there is no established connection between the SLE and the standard quantum master equations, the SLE is different from the stochastic Schr\"odinger equation (SSE) developed to mock the evolutions given by the quantum master equations \cite{Molmer:1993,Gisin:1992,Diosi:1998}. As stochastic equations based on pure state evolutions, they nevertheless share the same philosophy to perform an average over a large ensemble of initially identical subsystems to recover the statistical mixed state describing the subsystem\footnote{Because of the statistical nature of the bath-subsystem interactions, the subsystem must be described by a mixed state, which includes not only probabilistic information about the observable measurements but also about the state itself. The common tool to describe a mixed state is the density matrix operator.}. The expectation value of an observable operator $\hat{O}$ is then given by
\begin{equation}\label{MixedStochObservable}
\Big\langle\langle\psi(t)|\hat{O}|\psi(t)\rangle\Big\rangle_{\rm stat} = \lim_{n_{\rm stat}\rightarrow\infty}\,\frac{1}{n_{\rm stat}} \,\sum_{r=1}^{n_{\rm stat}} \langle\psi^{(r)}(t)|\hat{O}|\psi^{(r)}(t)\rangle\,,
\end{equation}
where the pure state $|\psi^{(r)}(t)\rangle$ is given by the $\mbox{r}^{\rm th}$ realization of the stochastic evolution. The numerical costs remain quite reasonable in comparison to the common density matrix approach which is highly expensive when the Hilbert space associated to the subsystem is large.\\

Before considering any actual application to phenomenology, some questions and issues remain to be explored about the solutions of the SLE and its thermal relaxation.

The study of its solutions without stochastic term has been carried out in many specific cases, either analytically \cite{Kan:1974,Dekker:1981,Haas:2013,Immele:1975,Hasse:1975,Skagerstam:1975,Brull:1984,Bassalo:2010,Zander:2013} or numerically \cite{Garas:2013,Sanin:2014,Immele:1975,Wells:1974,Falco:2009,Razavy:1978,Sanin:2007,Chou:2015}. Along these analysis, it has been advocated that the stationary eigenstates of $H_0$ are also stationary states of the equation \cite{Kan:1974,Skagerstam:1975}. This behavior is in contradiction with what is expected from damped quantum systems \cite{Lindblad:1976wv,Senitzky,Ford:1965,Caldeira:1983,Weiss:2012}. As an answer to this expectation, we will first show in Sec.~\ref{PolarPres} how a proper application of the Madelung/polar transformation of the wave function results in the damping of these states. The purely dissipative SLE has already been applied in quantum chemistry \cite{Garas:2013} and heavy ion scattering \cite{Sandulescu:1981,Hernandez:1981}.

However very few studies including the stochastic term have been carried out and these are moreover mere comparisons with the solutions of the HLE. Kostin \cite{Kostin:1972} first observed that for a free particle plane wave, the SLE and HLE lead to the same solution. Then, Messer \cite{Messer:1979} studied the evolution of a Gaussian wave packet in the free and harmonic cases. In the free case, he showed that the evolution differs from the HLE solution, highlighting that the SLE and HLE are not strictly equivalent. In his calculation, Messer used a white noise for the stochastic force $F_R(t)$ -- which is questionable -- and assumed that the SLE naturally leads to the thermal equilibrium predicted by statistical mechanics\footnote{Within these assumptions, the SLE has already been applied to atomic diffusion in solids \cite{Weiner:1974}.} (the Gibbs state). The latter is characterized by a Boltzmann distribution of the uncoupled subsystem energy states $\{E_n\}_{n=0,1...}$: 
\begin{eqnarray}\label{SubDistatEq}
p_n\propto\exp\left(\frac{-E_n}{k T_{\rm bath}}\right),
\end{eqnarray} 
where $k$ is the Boltzmann constant and $p_n$ the population (or ``weight") of the eigenstate of energy $E_n$, and is generally expected at the weak coupling limit for a quantum subsystem in interaction with a heat bath  \cite{Breuer:2002,Gardiner:2000,Biele:2014}. The so-called {\em weak coupling limit} or Brownian hierarchy is achieved when the relaxation time of the subsystem ($\sim 1/A$) is much larger than its natural period of oscillation and than the typical correlation time $\sigma$ of the microscopic interactions between the bath components and the subsystem. To our knowledge it has never been proven or tested that the SLE actually admits
such an asymptotic distribution in a dynamical manner. The main contribution of the present work is precisely to study the thermal relaxation given by the SLE with different 1D subsystems using either a white or a colored noise and thus to test Messer's assumption as a byproduct. To this end, in section \ref{QuantNoise}, we discuss different possible choices for the stochastic force $F_R(t)$ (which will be assumed to be a c-number). In section \ref{Section3Harmo}, we study the equilibration given by the SLE with a 1D harmonic potential $V_{\rm ext}=K\,x^2/2$. To do so, we first show analytically that the SLE brings a Gaussian wave packet to the Gibbs state if one uses a specific white noise. We generalize this result to other initial states through the numerical resolution of the SLE with the Crank-Nicolson scheme. Similarly, we then explore the equilibration resulting from a colored noise and demonstrate that it leads to the correct equilibration in the weak coupling limit. In section \ref{Section4Linear}, we extend the numerical simulations to a linear potential $V_{\rm ext}=K_l\,|x|/2$ to study the equilibration given by the SLE within a non-harmonic situation.

Though the thermal equilibrium of statistical mechanics is only expected at the weak coupling limit\footnote{Generally, a quantum subsystem in interaction with a heat bath is expected to reach a thermal equilibrium, where the components of its energy spectrum are shifted and broadened \cite{Breuer:2002,Gardiner:2000,Biele:2014}. These spectrum modifications become negligible at the weak coupling limit and one expects the thermal equilibrium of statistical mechanics. Note also that the equations of evolution, such as the quantum master equations and the HLE, are derived under the assumption of a weak coupling.} ($A\ll \{\omega_0, 1/\sigma\}$),
where $\omega_0$ is the characteristic frequency associated to 
the potential $V_{\rm ext}$\footnote{$\omega_0=\sqrt{\frac{K}{m}}$ for
the harmonic potential and $\omega_0=\sqrt[3]{\frac{K_l^2}{m \hbar}}$ for the linear	 potential.}, the intermediate ($A\lesssim \{\omega_0, 1/\sigma\} $) and strongly coupled regimes ($A\gtrsim \{\omega_0, 1/\sigma\}$) are also investigated.

Although we perform the numerical simulations with the dimensionless SLE for simplicity -- i.e. with natural units $\hbar=m=K=K_l=\omega_0=k=1$ and dimensionless variables $x$, $t$,\ldots\footnote{The dimensioned values of $x$, $t$, $A$, $F_R$ and $H_0$ can be obtained by multiplying our dimensionless values respectively by $\sqrt{\hbar/m\omega_0}$, $1/\omega_0$, $\omega_0$, $\sqrt{m\hbar\omega_0^3\,}$ and $\hbar\omega_0$, where $\omega_0$ is specific to each type 
of potential.} -- the analysis shows that the important dimensionless ratios governing the physics are $A/\omega_0$ and $k T_{\rm bath}/(\hbar \omega_0)$ for both potentials. In the main part of the text, 
it is then assumed, for the purpose of compactness, that $A$ and $T_{\rm bath}$ are respectively measured in units of $\omega_0$ and $\hbar \omega_0/k$ -- and thus correspond to these ratios --, while times are measured in units of $\omega_0^{-1}$ and energies in units of $\hbar \omega_0$. For all analytical calculations, we stick however to the International System of Units, for the sake of clarity.

Along this work, we will extract some effective temperature $T_{\rm sub}$ reached by the subsystem from the asymptotic weights $\{p_n\}_{n=0,1...}$, mostly by fitting a Boltzmann distribution $\propto\exp\left(-E_n/k T_{\rm sub}\right)$ to this equilibrium distribution. In some specific situations we will observe that $T_{\rm sub}$ differs from the bath temperature $T_{\rm bath}$ (which is concretely defined as the temperature entering the noise correlation); it is precisely one of the main goals of our study to identify under which conditions such a departure happens.

\section{A well defined prescription for the friction term to obtain eigenstates damping}\label{PolarPres}

Though theoretically the fluctuation and dissipation aspects cannot be dissociated, it appears that in some specific studies only the damping is considered \cite{Garas:2013,Chou:2015,Sandulescu:1981,Hernandez:1981}. Unfortunately, the dissipative part of the SLE (\ref{SLeq}) suffers from ambiguities and some prescription is required to obtain an unambiguous definition. Its main non-linear ingredient is the real phase $S(x,t)$, defined by the wave function decomposition  
\begin{equation}\label{MadTransfo}
\psi(x,t)=R(x,t)e^{iS(x,t)}\,,
\end{equation}
where $R(x,t)$ is the real amplitude. $S(x,t)$ is indeterminate at the wave function nodes and, in general, multivalued (defined modulo $2\pi$). 

In the literature \cite{Kan:1974,Garas:2013,Skagerstam:1975,Van:2004}, a common prescription adopted for real $\psi$ is to require the phase $S(\psi)$ to be zero (and thus continuous at the nodes of $\psi$) while $R(\psi)$ is taken as a real -- {\it positive or negative} -- function. This prescription has led to the conclusion that the stationary eigenstates of $H_0$ are also stationary states of the SLE, as the dissipation term identically vanishes. For the sake of describing time-dependent situations, a corresponding prescription has however to be adopted for any complex $\psi$ as well. It is easily seen that such an analytical continuation unavoidably has one branch cut in each half complex-plane, both of them starting from the origin. Taking for instance those branch cuts along the imaginary axis corresponds to 
\begin{equation}
\label{arctanPres}
S(\psi)=\arctan(\Im(\psi)/\Re(\psi)).
\end{equation}
with finite values of the friction potential in the SLE and then finite damping. Therefore, an infinitesimal modification of $\psi$ (associated to a slight deviation from real axis to complex plane) leads to a large variation of the associated damping of the quantum state, which is the sign of an ill-defined model.

\begin{figure}[!h]
 \centering
\begin{minipage}[b]{0.45\linewidth}
\iffigsdirectory
\includegraphics[height=45mm]{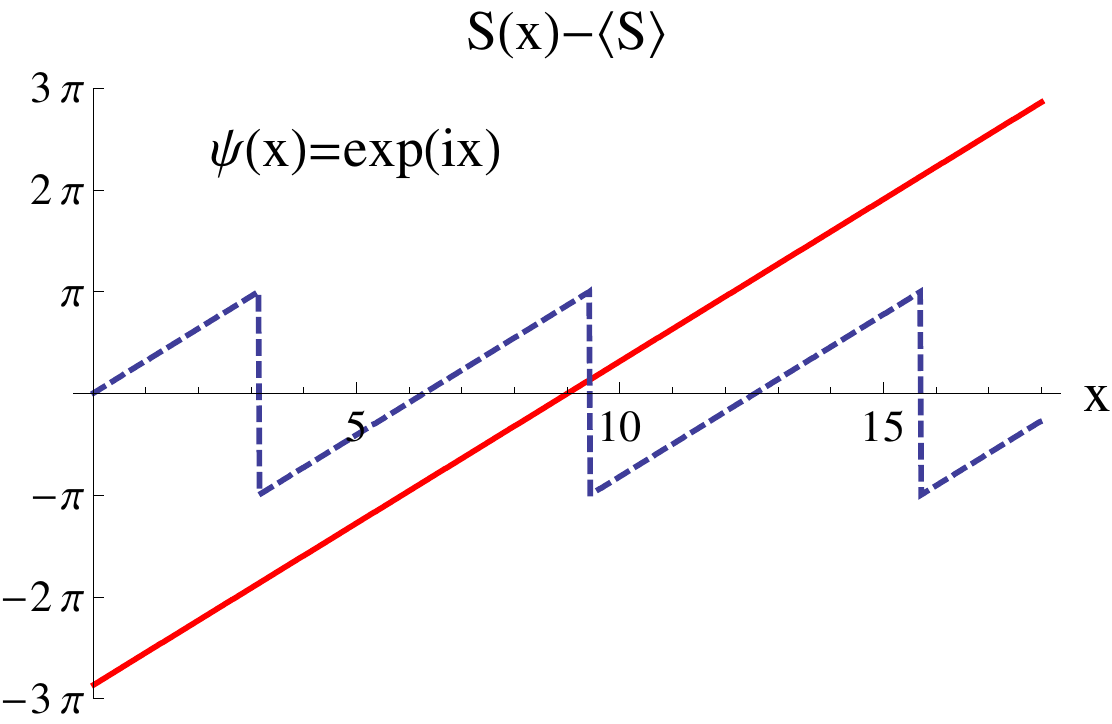}
\else
\includegraphics[height=45mm]{MonoChromandPhase.pdf}
\fi
 \caption{\label{fig: PlaneW}
   \small  The dissipative term $S(x)-\left<S\right>$ corresponding to the plane wave $\psi(x)=e^{ix}$ obtained with the argument (Arg) function (dashed line) and with the recursive method (solid line).}
\end{minipage}
\quad\,\,\,\,
\begin{minipage}[b]{0.45\linewidth}
\iffigsdirectory
\includegraphics[height=45mm]{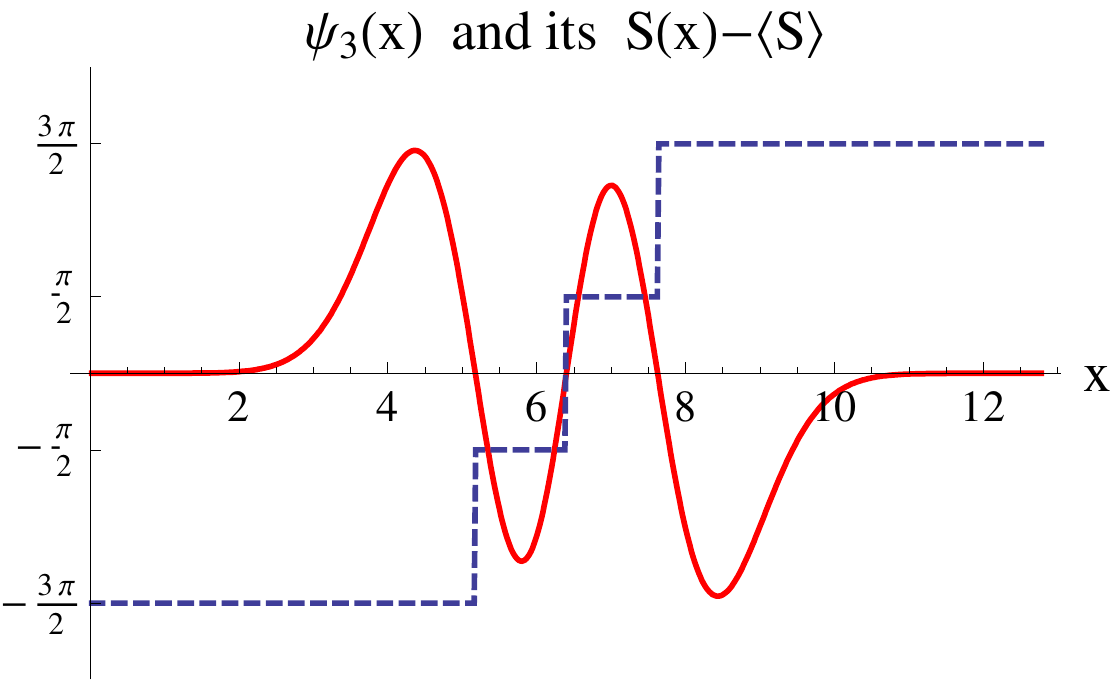}
\else
\includegraphics[height=45mm]{Psi3andPhase.pdf}
\fi
   \caption{\label{fig: Psi3}
   \small  The real $\psi_3$ harmonic eigenstate (solid line, magnified for the plot sake) and its corresponding phase $S(x)-\left<S\right>$ (dashed line) with the recursive method in the polar $\pi$ prescription.}
\end{minipage}
\end{figure}

We propose to use instead the ``polar" or ``Madelung" prescription, where one defines $R(x,t)$ in eq.~(\ref{MadTransfo}) as the module of the wave function, i.e. a real {\it positive} function. In practice, one could use the local argument of the wave function ${\rm Arg}(\psi)$ -- a well-defined function with a branch cut along the negative real axis\footnote{In many programming languages, such a choice of the argument of a complex number $z$ is given by the function atan2, i.e. ${\rm Arg}(z)={\rm atan2}(\Im(z),\Re(z))$.} -- to determine $S(x,t)$. However, the limitation of the Arg values to a $]-\pi,+\pi]$ interval (as illustrated in Fig.~\ref{fig: PlaneW}) would lead to discontinuities of the dissipative term with unphysical effects\footnote{The invariance under the multiplication of the wave function by a simple phase factor would be broken.}. To avoid these, 
we suggest to build the phase $S(x,t)$ on a spacial grid of step $dx$ following the recursive law
\begin{equation}\label{StepStep}
S(x+dx)=S(x)+dS(x)\quad \mbox{where}\quad dS(x)={\rm Arg}[\psi(x+dx)/\psi(x)],
\end{equation}
starting from an arbitrary space point of reference ``$0$". This leads to
\vspace{-1mm}
\begin{equation}\label{StepStep2}
S(j\times dx)=S(0)+\sum_{k=1}^j dS(k\times dx).
\end{equation}
The chosen value of the multivalued $S(0)$ is of no importance thanks to the regulator $-\left<S\right>$ and can therefore be taken to ${\rm Arg}(\psi(0))$. 

The polar prescription implies singular phase shifts $+\pi$ at the wave function nodes as shown for instance in Fig.~\ref{fig: Psi3}. Not only are these discontinuities theoretically allowed (thanks to the phase indeterminacy at the nodes), but they also have a convenient physical consequence: the eigenstates of $H_0$ are not stationary states of the SLE anymore. Indeed, for the excited eigenstates $\{\psi_n\}_{n\geq1}$ of $H_0$ the friction term becomes a step potential which generates correlations between eigenstates and results in damping. To show the latter assertions, let us assume that an initial wave function $\psi(0)$ is equal to an eigenstate $\psi_{m\geq1}$, i.e. $\psi(0)=\sum\,c_n(0) \,\psi_{n}$ with $c_n(0)=\delta_{nm}$. The SLE without the stochastic term yields,
\begin{eqnarray}\label{Correlationeq}\nonumber
\dot{c}_{n}&=&-\frac{i}{\hbar}\langle\psi_{n}|H_0|\psi\rangle-i A\left<\psi_{n}|(S-\left<S\right>)|\psi\right>\\
&=&-\frac{i}{\hbar} E_{n}\,c_{n} -i A \sum_k c_k\int (S-\left<S\right>) \psi_n^*\,\psi_k\, dx\,.
\end{eqnarray}
For symmetric potentials for instance, one can show that if $\psi_{k=m}$ has an odd (even) parity, then the integral is finite and thus the transition $c_m\rightarrow c_n$ is allowed at very small times for all $\psi_n$ with even (odd) parities. Moreover, the smaller the difference $|n-m|$, the larger the transition rate, which is consistent with the typical behavior of transition matrix elements entering e.g. the Fermi Golden Rule. Last but not least, the transition rate to $n=m-1$ is larger than to $n=m+1$, which is consistent with damping. At larger times, these transitions and the damping can be observed numerically (see for instance Fig.~\ref{fig: FrictionOnly}).\\ 
\begin{figure}[!h]\label{fig: FrictionOnly}
 \centerline{
 \iffigsdirectory
\includegraphics[height=50mm]{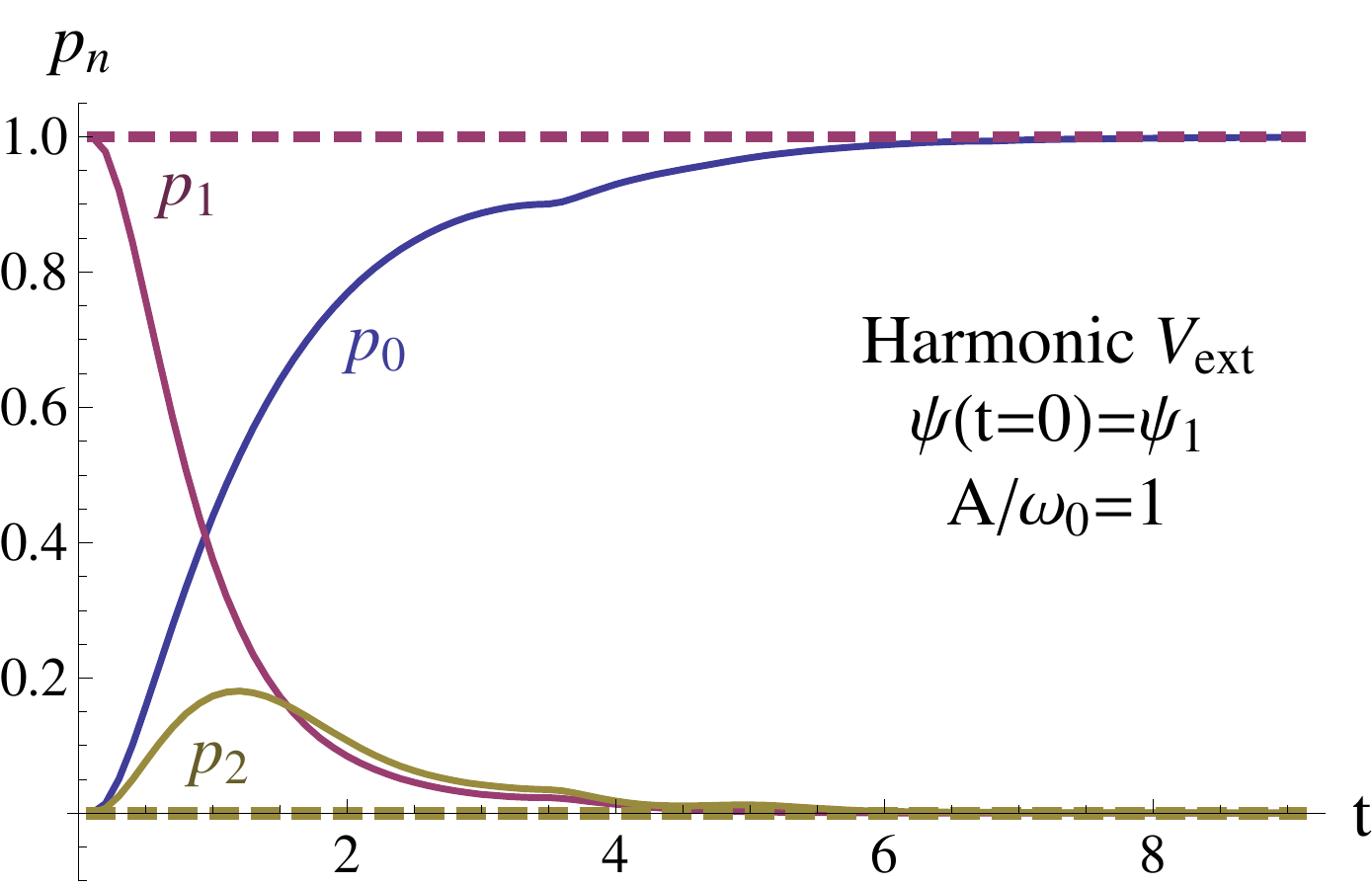} 
 \else
\includegraphics[height=50mm]{Psi1HarmoFrictionOnly2.pdf} 
 \fi
}
   \caption{\label{fig: FrictionOnly}
   \small  Evolution of the eigenstate weights $p_{n=0,1,2}=\vert\langle\psi_n\vert\psi (t)\rangle\vert^2$ with the dimensionless SLE without fluctuating term from an initial first excited state, with the ``polar" prescription (solid lines) and the ``arctan" prescription (dashed lines).}
\end{figure} 

Both the ``arctan" and ``polar" prescriptions are mathematically correct and the choice between them should be physically motivated. The stationarity of the $H_0$ eigenstates in the corresponding dissipative situation remains an open question within the open quantum system framework \cite{Tarasov:2011}. As illustrated in Fig.~\ref{fig: FrictionOnly}, the SLE can reproduce both situations, thanks to the two prescriptions presented in this section. From the perspective of the common quantum master equation \cite{Breuer:2002}, the Lamb shifted energy levels acquire finite lifetimes (finite widths), implying that the polar prescription is better suited for robust phenomenological studies. Let us finally stress that the choice of the prescription is of little importance when the fluctuations are considered, as they drive the state away from any given eigenstate.

\section{The possible noises for the stochastic term}\label{QuantNoise}\label{QNchoice}
We now focus on the stochastic aspect of the SLE (\ref{SLeq}). Similarly to other Langevin-like equations, the noise term $F_R(t)$ simulates the many collisions (or couplings) that the subsystem undergoes with the particles of the bath. It is generally taken as a homogeneous Gaussian random process, independent of the subsystem position, and described by its average and covariance function. The random direction of the many collisions always yields a zero average. The covariance function $C$ is usually given by a fluctuation-dissipation relation. The latter relates the noise covariance, the friction coefficient $A$ and the bath temperature $T_{\rm bath}$ in order to obtain a balance between the fluctuation and dissipation aspects. The balance is correct if the subsystem distribution at equilibrium is Boltzmannian $\propto\exp\left(-E_n/T_{\rm sub}\right)$ and if the temperature reached by the subsystem $T_{\rm sub}$ is equal to $T_{\rm bath}$. As the fluctuation-dissipation relation corresponding to the SLE has never been determined to our knowledge, our method will consist in borrowing such relation from another framework (discussion in this section) and then evaluating its consequences for the SLE (work done in next sections). As the SLE might be the counterpart of the HLE in the Schr\"odinger representation, the covariance functions derived within the HLE framework might be suited. Within this framework, the noise operator is built from the initial bath position and momentum operators whose non-commutative property leads to the main differences with the classical case.\\ 

Senitzky \cite{Senitzky} first proposed an HLE -- for a general bath linearly acting on a harmonic subsystem (with natural frequency $\omega_0$) -- where the noise operator is described by a white noise autocorrelation,
\begin{equation}\label{QNSen}
\left< \hat F_R(t) \hat F_R(t+\tau)\right>=C_{\rm Sen}(\tau)
\quad\text{with}\quad C_{\rm Sen}(\tau):=2mA\bigg[\frac{\hbar \omega_0}{2}+\frac{\hbar \omega_0}{\exp(\hbar \omega_0/kT_{\rm bath})-1} \bigg]\delta(\tau),
\end{equation}
where $\delta$ is the Dirac distribution. As in the classical case, a white noise covariance means that there is no correlation between the collisions, which results in a Markovian process. This covariance has been used by Messer \cite{Messer:1979} in its analytic comparison of the HLE and SLE solutions. The first term of the RHS bracket in the definition of $C_{\rm Sen}$ corresponds to the zero point fluctuations of the subsystem. This term is required within the HLE framework for the canonical commutations to hold at $T_{\rm bath}=0$, as shown by equation (52) in \cite{Senitzky}. However, within the SL framework, the zero point fluctuations appear naturally in the wave function, so that they do not need to be included in the noise operator for the canonical commutations to hold. Therefore, this term becomes unnecessary and the white quantum noise autocorrelation $C_{\rm white}$ will be defined as 
\begin{eqnarray}
\label{QNRK}
\label{QNRKB}
C_{\rm white}(\tau):=B\,\delta(\tau)
\quad\text{with}\quad B:=2mA\,E_0\bigg[\coth\bigg(\frac{E_0}{kT_{\rm bath}}\bigg)-1\bigg]\,.
\end{eqnarray}
where $E_0=\hbar \omega_0/2$, the zero point energy. 
In Sec.~\ref{Section3Harmo}, we will show that the fluctuation-dissipation relation (\ref{QNRKB}) indeed allows to reach asymptotically a thermal distribution of states when one uses a white noise and a harmonic potential. 

However, Li et al.~\cite{Xi:1995} pointed out an important weakness in the derivation of (\ref{QNSen}). They also claimed that the colored quantum noise 
\begin{eqnarray}\label{QNFord}
\left<\hat F_R(t) \hat F_R(t+\tau)\right>=\frac{m}{\pi}\int_0^\infty \hbar \omega \bigg[ \coth\bigg(\frac{\hbar \omega}{2kT_{\rm bath}}\bigg) \cos(\omega \tau)+i\sin(\omega \tau) \bigg] A\, d\omega\,,
\end{eqnarray}
first derived by Ford et al.~\cite{Ford:1965}, is the only one able to drive a general subsystem to the correct thermal equilibrium via the HLE. For now, the latter assertion has only been demonstrated in a limited form for harmonic and nearly harmonic oscillators \cite{Gardiner:2000,Benguria:1981}. 

Actually, in order to get rid of the contribution from the bath zero point fluctuations -- which was first judged physically unjustified --, Ford et al.~first derived a colored quantum noise under the form of the normal product
\begin{equation}\label{QNFordNormProd}
\left<N[\hat F_R(t) \hat F_R(t+\tau)]\right>=
C_{\rm colored}(\tau):=\frac{2mA}{\pi}\int_0^\infty \frac{\hbar \omega}{\exp(\hbar \omega/kT_{\rm bath})-1} \cos(\omega \tau)\, d\omega.
\end{equation}
Both colored noises lead to a non-Markovian process even if the friction is memory-less. As pointed out by Gardiner \cite{Gardiner:2000}, the correct choice of spectrum depends on what is actually measured to find it: e.g.~in absorption measurements one gets the black body radiation Planck spectrum corresponding to (\ref{QNFordNormProd}), whereas in Josephson junction noise current measurements \cite{Koch:1981} one gets the linearly rising spectrum at high frequencies corresponding to (\ref{QNFord}). 
Within the SSE framework, the choice between the different noises is also intensively discussed, for instance to obtain the correct thermal equilibrium of a non-Markovian master equation \cite{Biele:2014} or the correct positivity property for the Bloch--Redfield master equation \cite{Whitney:2008}. \\ 

In general, the practical application of the HLE is limited by its non-commutating operator nature. Although questionable \cite{Hanggi:2005}, a common approximation \cite{Koch:1980,Sebastian:1981,Schmid:1982,Metiu:1984,Eckern:1990,Banerjee:2003,Banerjee:2004} is to abandon its operator character and to replace the non-commutating q-number noise by a c-number noise while taking the same power spectrum. One then obtains a quasiclassical Langevin equation which leads to a reasonable description for systems which are nearly harmonic and to possible violations of the Heisenberg relations. Within the SLE, we only need to assume that the noise operator can be taken as a commutating c-number. The latter assumption was actually already implied in Kostin's derivation of the SL random potential \cite{Kostin:1972} and does not lead to a violation of the Heisenberg relations \cite{Dekker:1981,Haas:2013,Sanin:2014}. 

In the present paper, we adopt the same assumption: 
$F_R$ will be considered as Gaussian stochastic c-numbers of zero average while the autocorrelation $C$ will be taken according either to definition (\ref{QNRK}) (white noise) or definition (\ref{QNFordNormProd}) 
(colored noise). We will then test the ability of the SLE
to bring a subsystem to thermal equilibrium.
Whereas the white noise (\ref{QNRK}) leads to uncorrelated stochastic forces, the colored noise (\ref{QNFordNormProd}) contains a strong temperature dependence of their correlation time. The latter becomes large at low temperatures ($\propto 1/T_{\rm bath}$) and the Brownian hierarchy/weak coupling limit could be broken in such a situation, when $A\gtrsim T_{\rm bath}$ (see introduction). 
In \ref{appendixNoise}, we describe the algorithm used to build these correlated forces numerically. The colored noise (\ref{QNFord}) would have also been pertinent because of its direct connection with the quantum dissipation-fluctuation theorem, but it leads to additional complications (imaginary part) and model dependences. Indeed, the noise correlation (\ref{QNFord}) strongly depends on the value of its high frequency cut-off -- which evaluation is specific to each system -- and on the choice of the cut-off shape (Lorentzian, exponential, sharp...).

\section{Equilibration with a harmonic oscillator}\label{Section3Harmo}

In this section, we study the thermal relaxation given by the SLE (\ref{SLeq}) with the harmonic 1D potential $V_{\rm ext}=m\omega_0^2\,x^2/2$ and with the white (\ref{QNRK}) or colored (\ref{QNFordNormProd}) noise. The harmonic oscillator is a well-known basis to study the properties of an open quantum system formalism \cite{Weiss:2012}.

\subsection{Wave function evolution during one stochastic realization}
\label{AnalHarmo}\label{Scenario}

We first focus on the evolution of the wave function during one stochastic realization of the SLE. Analytically, one can study the evolution of a general Gaussian wave packet {\em Ansatz}
\begin{eqnarray}\label{PsiGaussWa}
\psi_A(x,t)=e^{\frac{i}{\hbar}\big(\alpha(t)(x-x_{\rm cl}(t))^2+p_{\rm cl}(t)(x-x_{\rm cl}(t))+\gamma(t)\big)},
\end{eqnarray}
where $\alpha(t)$ is a complex number related to the wave packet width (${\rm Im}(\alpha)>0$ at all times), $\gamma(t)$ a complex phase whose imaginary part plays the role of a normalizing factor, and $x_{\rm cl}$ and $p_{\rm cl}$ are 
centroids (and expectation values) in direct and momentum space. As shown in \ref{appendixGaussianevolve}, inserting (\ref{PsiGaussWa}) in the SLE leads to ordinary differential equations for $\alpha,\, x_{\rm cl},\, p_{\rm cl}$ and $\gamma$, including:
\begin{eqnarray}\label{alphaeq}
\dot{\alpha}+A \Re(\alpha)+\frac{2}{m}\alpha^2+\frac{m\omega_0^2}{2}=0\,
\end{eqnarray}
and
\begin{eqnarray}\label{xpPFD}
\dot{p}_{\rm cl}=-m\omega_0^2\,x_{\rm cl}-Ap_{\rm cl}+F_R\,, \qquad \dot{x}_{\rm cl}=\frac{p_{\rm cl}}{m}
\end{eqnarray}
In \ref{appendixGaussianevolve}, it is also demonstrated that the solution of eq.~(\ref{alphaeq}) starting from any $\Im(\alpha(t=0))>0$ tends asymptotically to $\alpha(t\rightarrow \infty)=i m\omega_0/2$, which corresponds to the width of the ground state $a=\sqrt{\hbar/m\omega_0}\,$. After some initial relaxation, the general solution (\ref{PsiGaussWa}) from any initial Gaussian state is thus the ground state displaced in direct and momentum space with a stochastic trajectory obeying the classical equations of motion (\ref{xpPFD}). 

\begin{figure}[!h]\label{fig: OneRealMod}
 \centerline{
 \iffigsdirectory
\includegraphics[height=25mm]{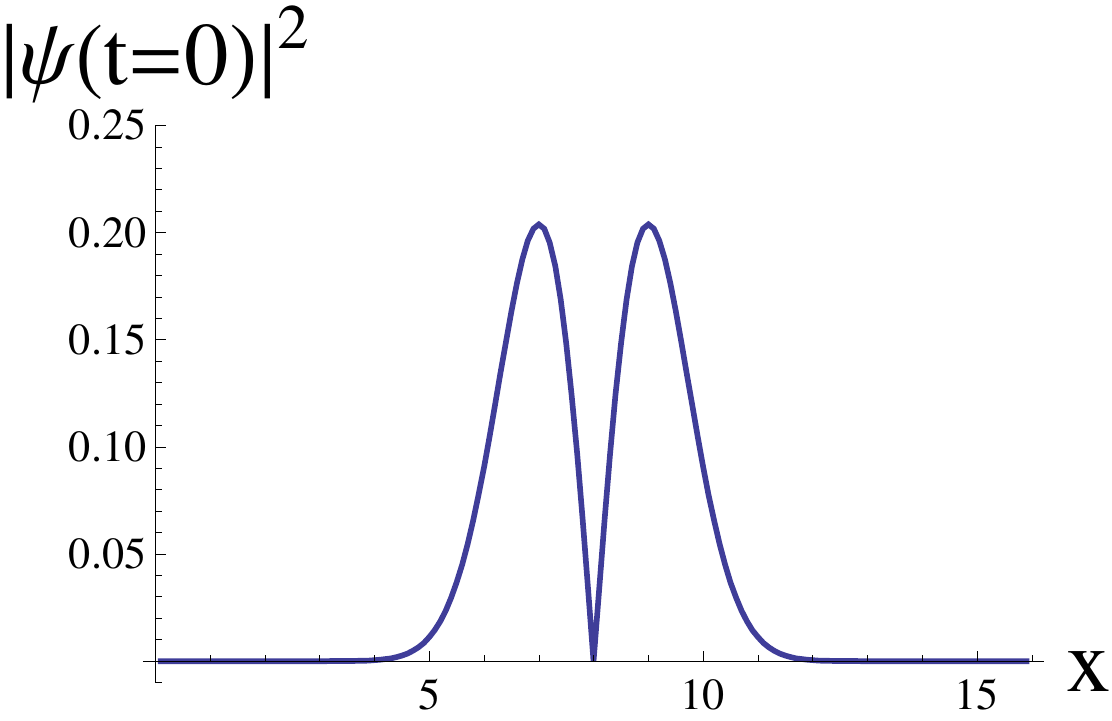}
\includegraphics[height=25mm]{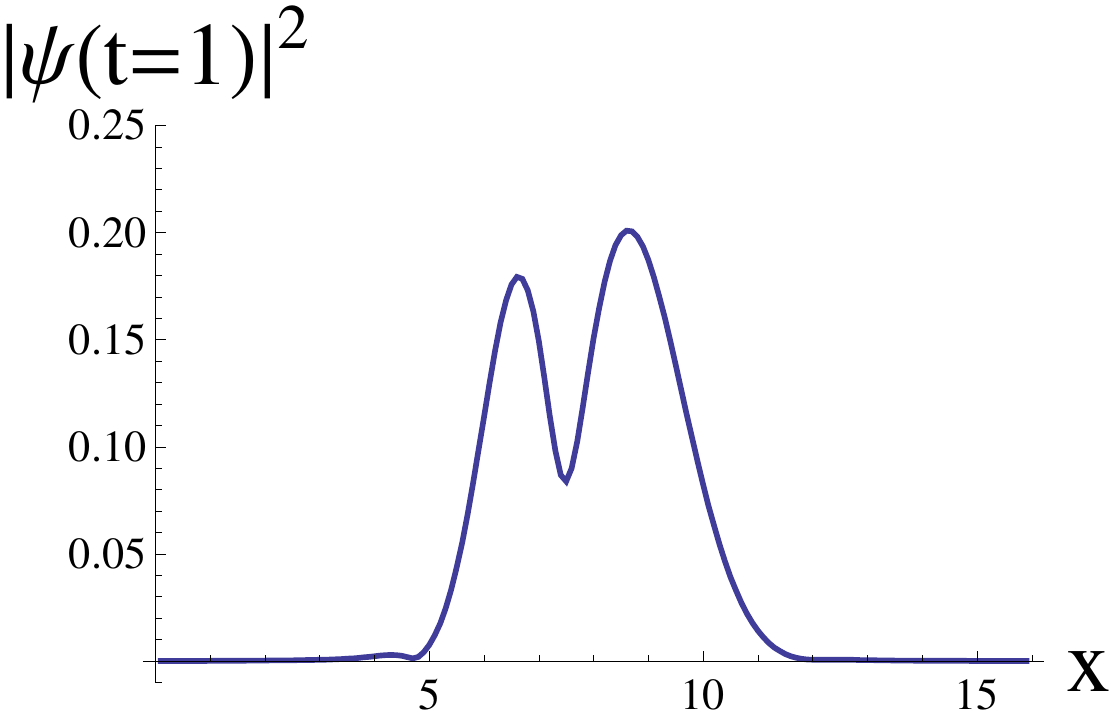}
\includegraphics[height=25mm]{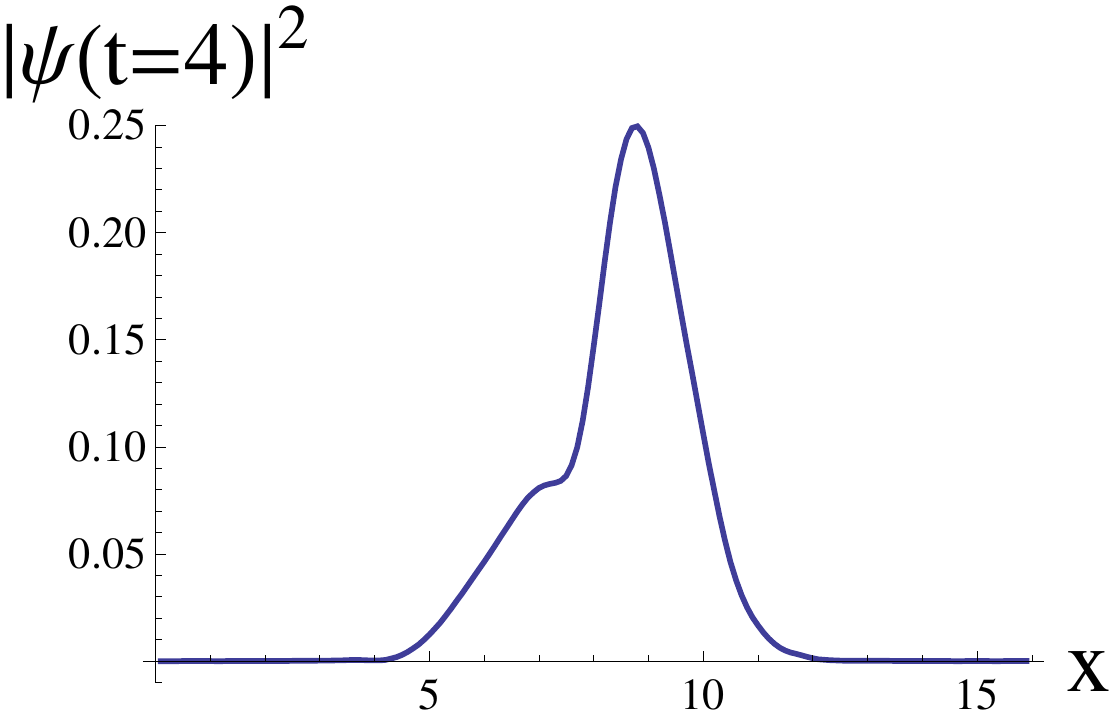}
\includegraphics[height=25mm]{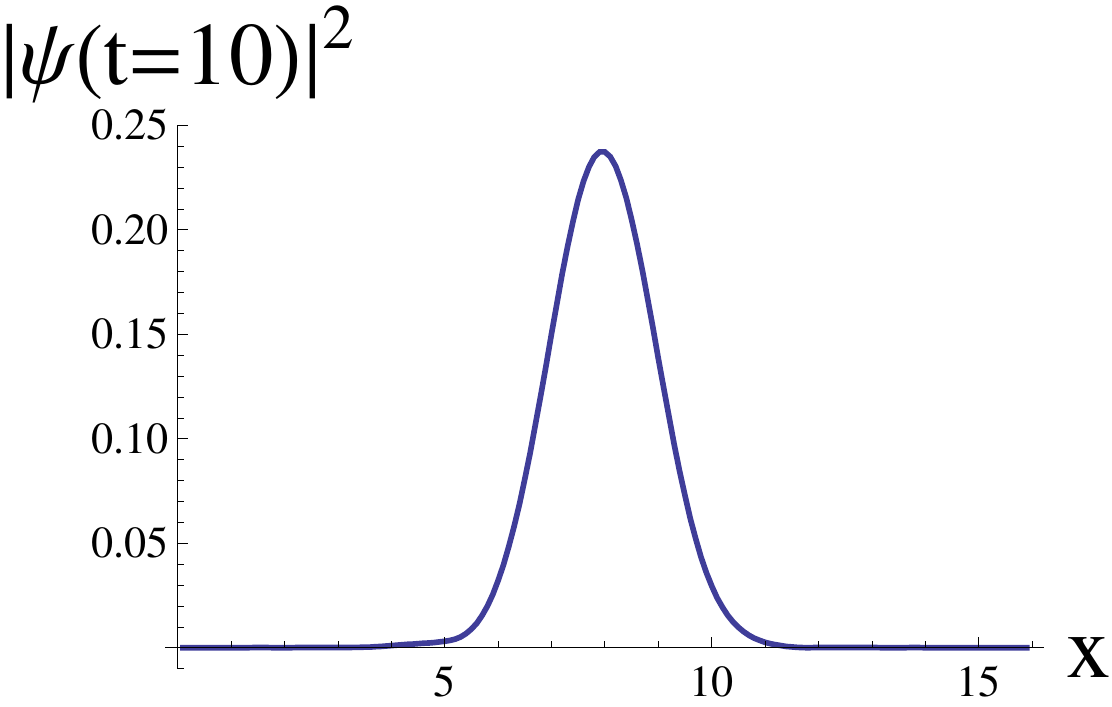} 
 \else
\includegraphics[height=25mm]{OneRealEvolMod1.pdf}
\includegraphics[height=25mm]{OneRealEvolMod2.pdf}
\includegraphics[height=25mm]{OneRealEvolMod3.pdf}
\includegraphics[height=25mm]{OneRealEvolMod4.pdf} 
 \fi}
   \caption{\label{fig: OneRealMod}
   \small  Typical wave function shape/module evolution toward the ground state shape/module during one noise realization.}
\end{figure} 
\begin{figure}[!h]\label{fig: OneRealPhase}
 \centerline{
 \iffigsdirectory
\includegraphics[height=24mm]{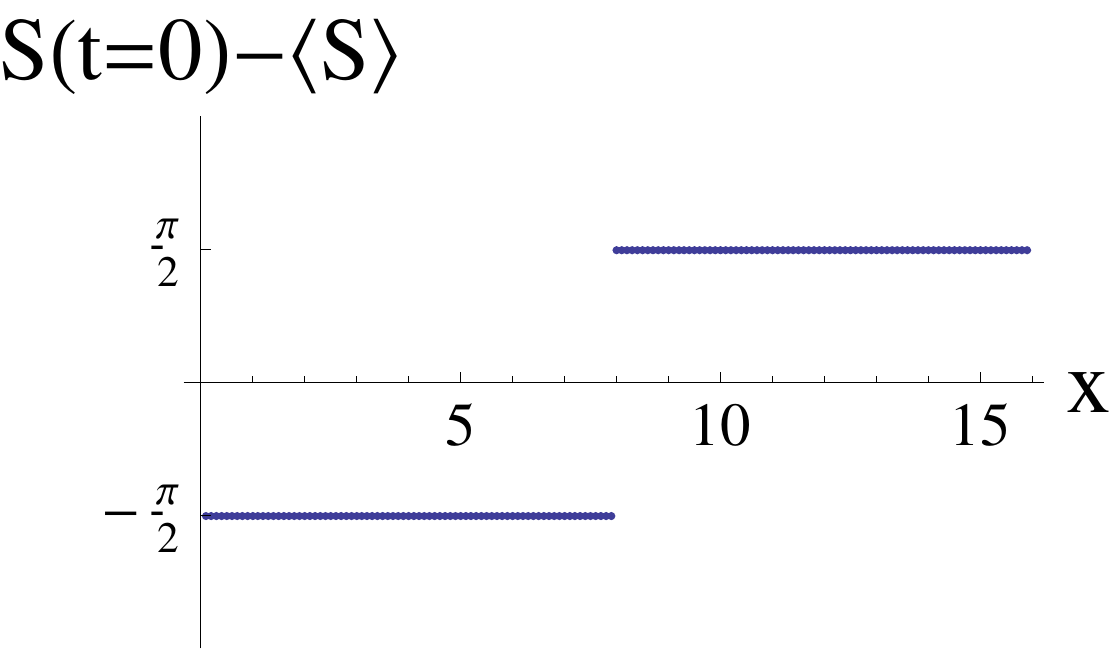}
\includegraphics[height=24mm]{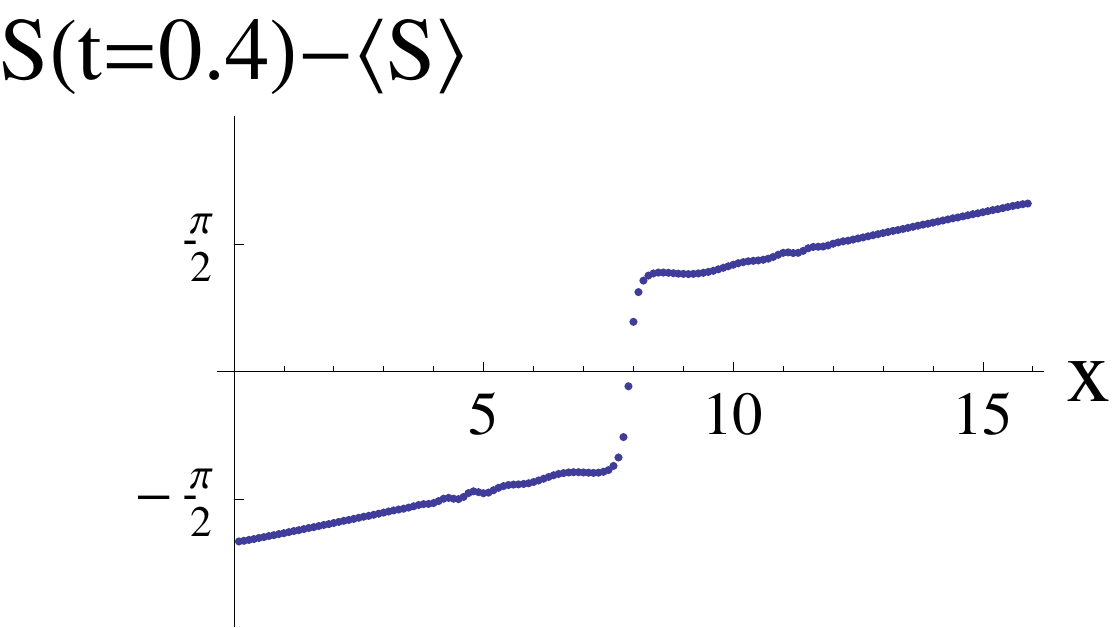}
\includegraphics[height=24mm]{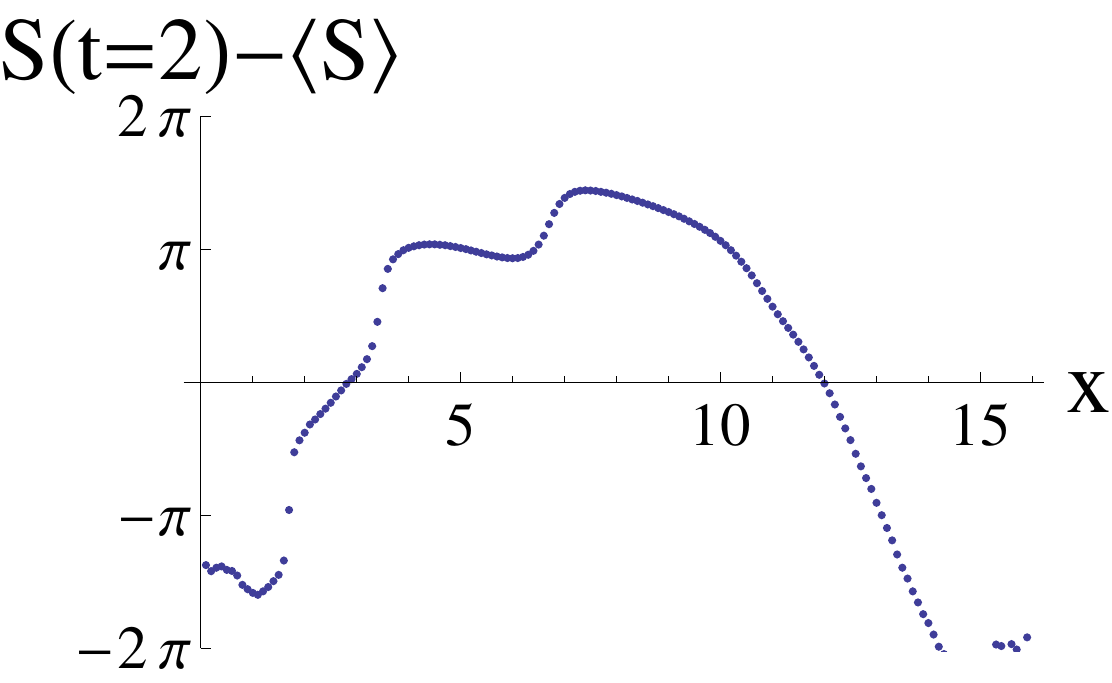}
\includegraphics[height=24mm]{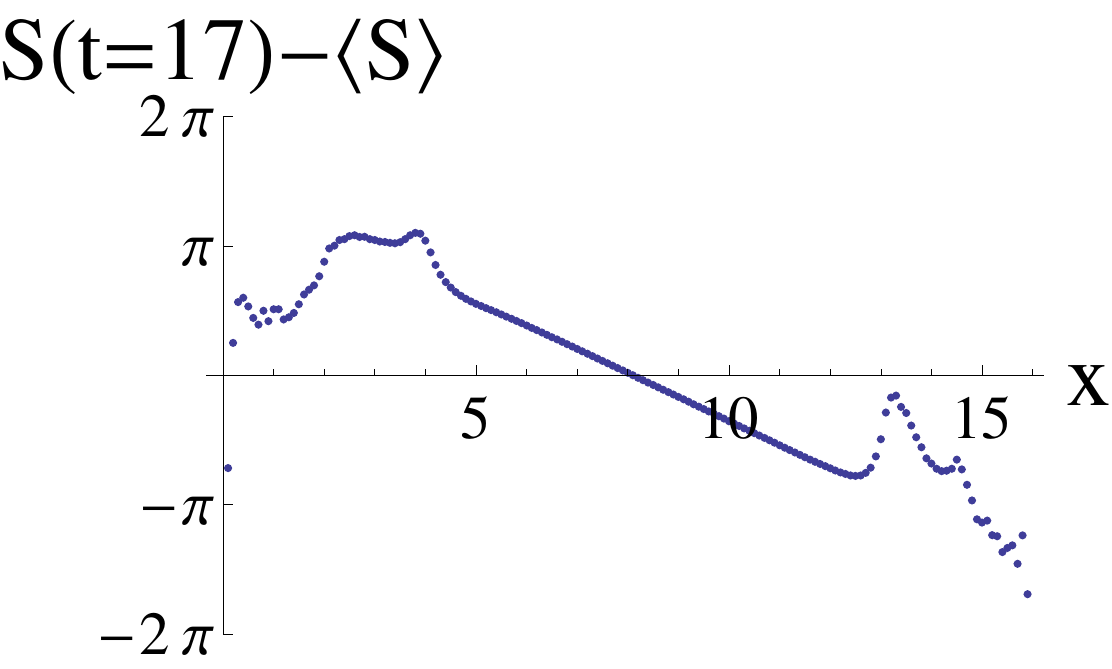} 
 \else
\includegraphics[height=24mm]{OneRealEvolPhase1.pdf}
\includegraphics[height=24mm]{OneRealEvolPhase2.pdf}
\includegraphics[height=24mm]{OneRealEvolPhase3.pdf}
\includegraphics[height=24mm]{OneRealEvolPhase4.pdf} 
 \fi}
   \caption{\label{fig: OneRealPhase}
   \small  Typical wave function phase evolution toward linearity (where the wave function takes significant values) during one noise realization.}
\end{figure} 

This result can be generalized to other initial states through a numerical resolution of the dimensionless SLE with the Crank--Nicolson scheme, which requires solving the equation
\begin{equation}
\left(1+\frac{i \hat{H}(t)\Delta t}{2 \hbar}\right)\psi(t+\Delta t)=
\left(1-\frac{i \hat{H}(t)\Delta t}{2 \hbar}\right)\psi(t)
\label{eq:crank-Nicolson}
\end{equation}
with respect to $\psi(t+\Delta t)$  at each time step $\Delta t$. For this purpose, the wave function has first been discretized on a spatial grid, with the grid size chosen large enough to avoid spurious reflections and the grid spacing much smaller then the typical inverse wave number. Next, the Thomas algorithm was applied for the inversion 
of equation (\ref{eq:crank-Nicolson}). Note that the stochastic force
$F_R$ was considered to be constant on each time step. This leads to an effective autocorrelation time $\sim \Delta t$ in the white noise
case, which has however shown to have no practical consequence. Finally, note that the numerical cost when considering ensemble averaged observables is proportional to the space-time grid and to the number of realizations, i.e.~to $n_{\rm space}\times n_{\rm time}\times n_{\rm stat}$ (where typically $n_{\rm space}$ is of the order of the hundreds, $n_{\rm time}$ and $n_{\rm stat}$ of the thousands).\\

As illustrated in Fig.~\ref{fig: OneRealMod} and \ref{fig: OneRealPhase}, we first confirm from observations that after some initial relaxation, the general solution from any initial state is the ground state displaced in direct and momentum space with a stochastic trajectory. Indeed, a common wave function evolution pattern emerges during a noise realization. First, the shape of the wave function evolves toward the ground state shape. In parallel, if one starts from an initial excited eigenstate, the phase ``breaks" at the nodes and evolves toward a linear phase in the region where the wave function takes non negligible values (see at $t=17$ in Fig.~\ref{fig: OneRealPhase} for instance). In parallel and until the end of the evolution, the centroid oscillates around the potential minimum following a stochastic trajectory along the space axis. Some discrepancies to this pattern, coming from numerical instabilities, appear when $A\ll T_{\rm bath}$ and when $T_{\rm bath}\gg 1$.

\subsection{Evolutions of mixed state observables: energy and populations}\label{HarmoObs}

We now focus on the evolutions of mixed state observables as given by the statistical relation (\ref{MixedStochObservable}). To do so, we perform the numerical simulations\footnote{The grid steps are taken as $\Delta x=0.1$ and $\Delta t=0.01$. The covariance parameter of the white noise is taken to $\sigma=0.03$; see remark after equation (\ref{PSw}).} in a bath at temperature $T_{\rm bath}=1$ and with the white noise (\ref{QNRK}) for illustration. 

\begin{figure}[!h]\label{fig: EnergyAv}
 \centerline{
 \iffigsdirectory
 \includegraphics[height=50mm]{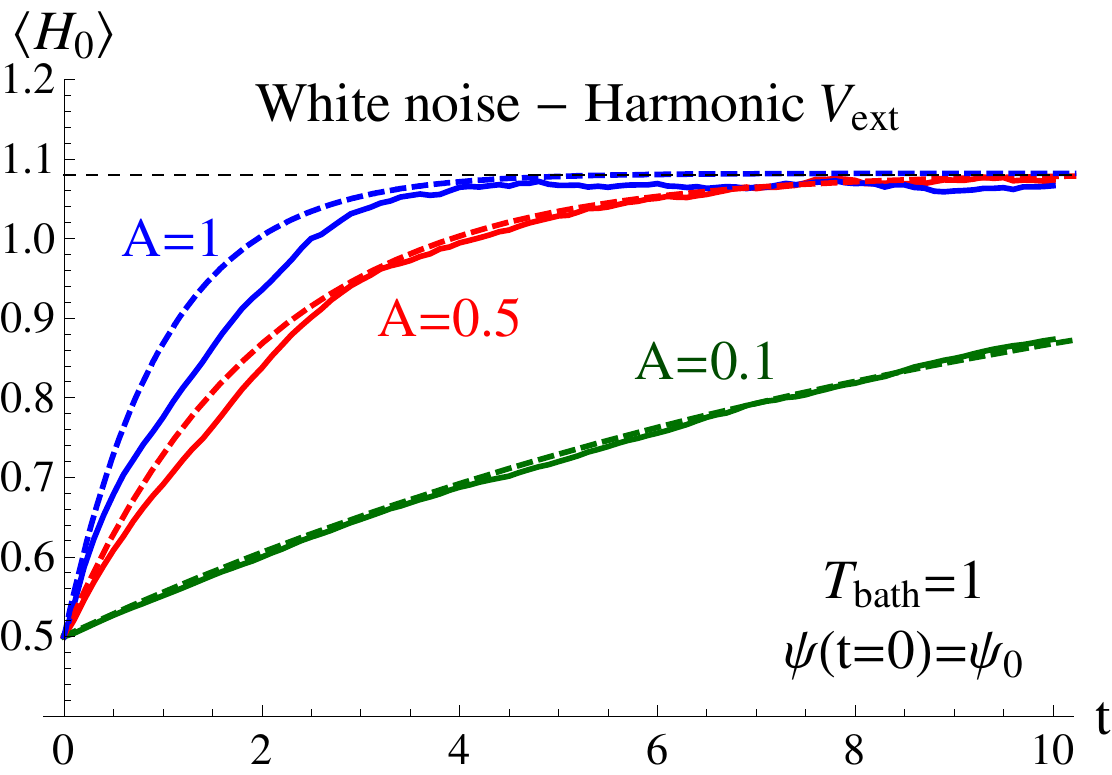}
 \else
 \includegraphics[height=50mm]{AverageEnergyEvolHarmoWhite.pdf}
 \fi
}
   \caption{\label{fig: EnergyAv}
   \small   {\it Solid curves}: Numerical average energy  
  ($\langle H_0 \rangle$) evolutions for different values of the friction coefficient $A$. {\it Dashed horizontal line}: Corresponding HLE theoretical asymptotic value given by the exact relation (\ref{ExactAsymAvEn}). {\it Dashed curves}: Corresponding theoretical evolutions given by relation (\ref{SenEnerEvol}).}
\end{figure} 

We first study the evolution of the average energy $\big\langle\langle H_0 \rangle\big\rangle_{\rm stat}$, which will be written $\langle H_0 \rangle$ for simplification, starting from the initial ground state $\psi_0$. Three average energy evolutions with friction coefficients corresponding to weak ($A=0.1$), intermediate ($A=0.5$) and strong ($A=1$) couplings 
(see introduction) are shown in Fig.~\ref{fig: EnergyAv}.

The average energy evolution rate predicted by Senitzky \cite{Senitzky} within the HLE framework,
\begin{equation}\label{SenEnerEvol}
\langle H_0\rangle(t) =E_0\,e^{-A t}+\langle H_0\rangle (t\rightarrow\infty)\Big(1-e^{-A t}\Big)\,,
\end{equation}
fits well our numerical evolution in the weak coupling case (where Senitzky's HLE actually applies) as shown in Fig.~\ref{fig: EnergyAv}. Furthermore, the theoretical asymptotic value for a quantum harmonic oscillator in thermal equilibrium is given by,
\begin{equation}\label{ExactAsymAvEn}
\langle H_0\rangle (t\rightarrow\infty) = E_0 \coth\left(\frac{E_0}{T_{\rm bath}}\right),
\end{equation}
and corresponds to our value $\langle H_0\rangle(t\rightarrow\infty)\simeq 1.07\,$ when $T_{\rm bath}\simeq 1$ for any coupling.

\begin{figure}[!h]\label{fig: HarmWeights}
 \centerline{
 \iffigsdirectory
 \includegraphics[height=50mm]{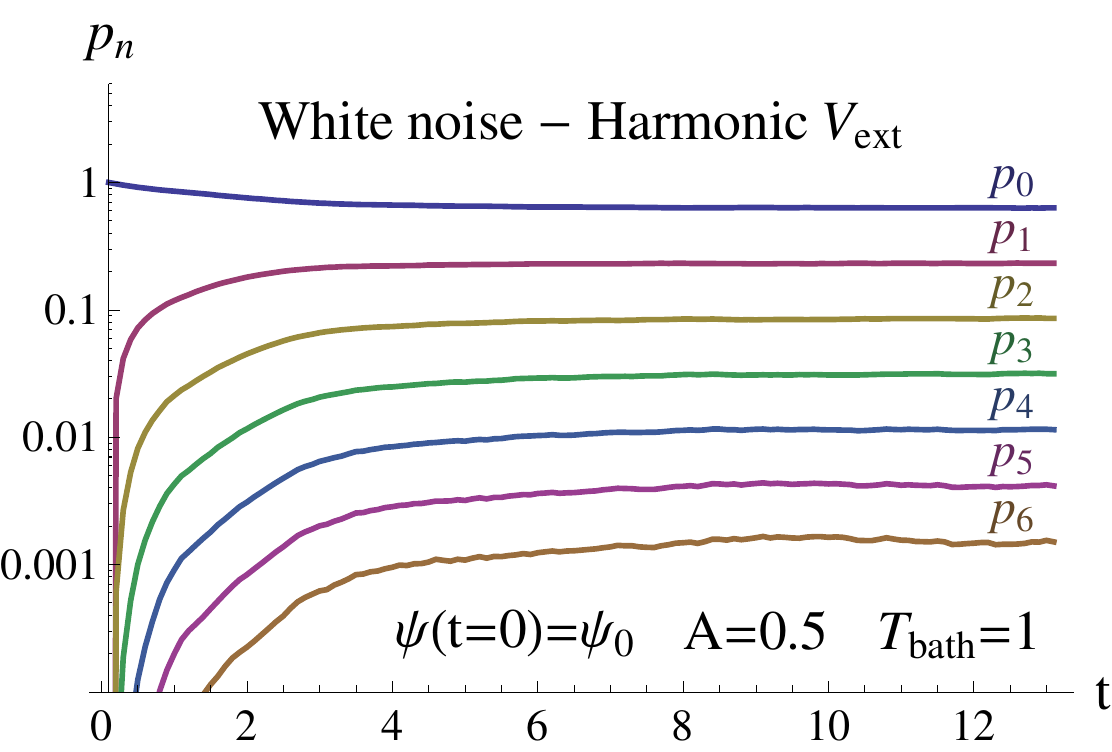}
 \includegraphics[height=50mm]{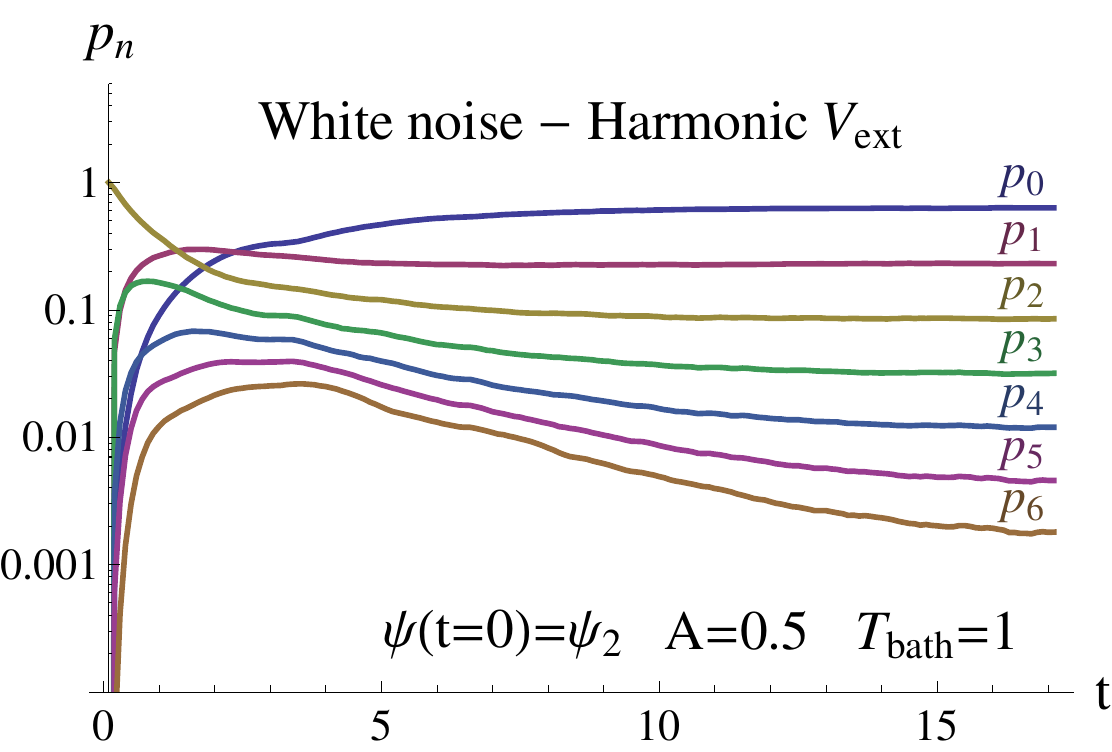}
 \else
\includegraphics[height=50mm]{Psi0HarmoWeightsEvol05_Log.pdf}
\includegraphics[height=50mm]{Psi2HarmoWeightsEvol05_Log.pdf} 
 \fi
}
   \caption{\label{fig: HarmWeights}
   \small  Evolutions of the eigenstate weights $p_{n=0,...6}(t)$ from the initial ground state ({\it left}) and $2^{\rm nd}$ excited state ({\it right}) for a friction coefficient corresponding to an intermediate coupling.}
\end{figure}

The second interesting observable is the distribution of the eigenstate weights (populations) $p_{n}(t)$, as given by relation (\ref{MixedStochObservable}) with the projection operator $\hat{O}_n=|\psi_n\rangle\langle\psi_n|$. As shown in Fig.~\ref{fig: HarmWeights}, their evolutions during the transient phase follow the general expectation of open quantum systems: the main transitions occur between neighboring energy levels. Moreover, they lead to a reshuffling of the weights, such as $p_n>p_{n+1}$, reached after a lapse of time proportional to the relaxation time $1/A$. 

\subsection{Equilibration with a white noise}\label{Section3HarmoWhite}

Here we show that the equilibrium distribution of the eigenstate weights is a Boltzmann distribution $\propto e^{-E_n/T_{\rm sub}}$ with $T_{\rm sub}=T_{\rm bath}$ provided that the fluctuation-dissipa-tion relation (\ref{QNRKB}) is satisfied if one uses a white noise. In other terms, the subsystem equilibrates with the bath if (\ref{QNRKB}) is satisfied when one uses a white noise. We first focus on the analytic solution with initial Gaussian wave packets and then show through numerical simulations that these results are universal, i.e.~independent of the chosen initial state, friction coefficient and temperature.

\subsubsection{With Gaussian initial wave packets}\label{AnalHarmoAsympt}

We showed in Sec.~\ref{AnalHarmo} that any Gaussian initial wave packet reduces asymptotically to a coherent state with a width  $a=\sqrt{\frac{\hbar}{m \omega_0}}$ corresponding to the ground state, i.e.
\begin{equation}\label{psiGauss}
\psi(t\gg A^{-1}) \propto e^{-\frac{(x-x_{\rm cl}(t))^2}{2 a^2}+i \frac{p_{\rm cl}(t) x}{\hbar}}\,\,.
\end{equation}
We would like to evaluate the weight of the different $H_0$ eigenstates
$\psi_n$ in this asymptotic $\psi$. For this purpose,
we first reformulate (\ref{psiGauss}) as
\begin{equation}
\psi \,\propto\, e^{-\frac{\xi^2}{2} +  2\mu \xi - \frac{(x_{\rm cl}/a)^2}{2}}\,,
\end{equation}
where we have set  $\xi=\frac{x}{a}$ and $\mu=\frac{\frac{x_{\rm cl}}{a}+ i \frac{p_{\rm cl} a}{\hbar}}{2}$. Using the identity
\begin{equation}
e^{2\mu \xi -\mu^2}= \sum_{n=0}^{+\infty} \frac{\mu^n}{n!} H_n(\xi)\,,
\end{equation}
where $\{H_n\}_{n=0,1...}$ are the Hermite polynomials, 
as well as the expression for the eigenstates
\begin{equation}
\psi_n= \frac{H_n(\xi) e^{-\frac{\xi^2}{2}}}{\sqrt{2^n n! \sqrt{\pi}}}\,,
\end{equation}
yields
\begin{eqnarray}
\psi \,\propto\, e^{- \left(\frac{x_{\rm cl}}{2 a}\right)^2 - \left(\frac{p_{\rm cl} a}{2\hbar}\right)^2 + i \frac{p_{\rm cl}x_{\rm cl}}{2 \hbar}} \sum_{n=0}^{+\infty} \frac{\sqrt{2^n}\mu^n}{\sqrt{n!}} \psi_n(\xi).
\end{eqnarray}
We thus deduce that the eigenstate weight $p_n(x_{\rm cl},p_{\rm cl})$ for a given realization of the stochastic noise is given by
\begin{equation}\label{Weightsxclpcl}
p_n(x_{\rm cl},p_{\rm cl})\,\propto\,  \frac{2^n|\mu|^{2n}}{n!}\,\, e^{-\frac{1}{2}\left(\frac{x_{\rm cl}}{a}\right)^2 -\frac{1}{2} \left(\frac{p_{\rm cl} a}{\hbar}\right)^2}\,\,
\propto\,\, \frac{
\left(\frac{1}{2}(\frac{x_{\rm cl}}{a})^2 + \frac{1}{2}(\frac{p_{\rm cl} a}{\hbar})^2\right)^n}{n!}\,\, e^{-\frac{1}{2}\left(\frac{x_{\rm cl}}{a}\right)^2 -\frac{1}{2} \left(\frac{p_{\rm cl} a}{\hbar}\right)^2}
\end{equation}
and one has exactly $\sum p_n=1$. In Sec.~\ref{AnalHarmo}, we have shown that the position $x_{\rm cl}$ and momentum $p_{\rm cl}$ centroids satisfy the classical stochastic equation of motion (\ref{xpPFD}). When the stochastic force autocorrelation is of the form $\left<F_R(t)F_R(t+\tau)\right>=B\,\delta(\tau)$ (white noise) it is known that the distribution of the trajectories ($x_{\rm cl},\,p_{\rm cl}$) is 
\begin{equation}
W(x_{\rm cl},p_{\rm cl}) \propto e^{-\frac{\frac{m\omega_0^2x_{\rm cl}^2}{2}+\frac{p_{\rm cl}^2}{2m}}{k T_{\rm eff}}},
\end{equation}
where $T_{\rm eff}:=\frac{B}{2 m A}$. The eigenstate weight, averaged over the fluctuations, will then be given by
\begin{equation}\label{WeigthIntegral}
p_n=\int W(x_{\rm cl},p_{\rm cl}) p_n(x_{\rm cl},p_{\rm cl}) dx_{\rm cl} dp_{\rm cl}\,.
\end{equation}
To determine (\ref{WeigthIntegral}), we use the relation
\begin{equation}
p_n(x_{\rm cl},p_{\rm cl})=\frac{(-1)^n}{n!}\frac{\partial^n}{\partial \eta^n} \left.e^{-\eta\left(\frac{(x_{\rm cl}/a)^2}{2}+\frac{(p_{\rm cl} a)^2}{2}\right)}
\right|_{\eta=1}\,.
\end{equation} 
After some trivial integration on $x_{\rm cl}$ and $p_{\rm cl}$, one gets that
\begin{equation}\label{IntWeights}
\int W(x_{\rm cl},p_{\rm cl}) e^{-\eta\left(\frac{(x_{\rm cl}/a)^2}{2}+\frac{(p_{\rm cl} a)^2}{2}\right)} dx_{\rm cl} dp_{\rm cl}=
\frac{\frac{\hbar\omega_0}{k T_{\rm eff}}}{\eta+\frac{\hbar\omega_0}{k T_{\rm eff}}}\,,
\end{equation} 
where the numerators guarantees that for $\eta=0$, one has $\int W(x_{\rm cl},p_{\rm cl})  dx_{\rm cl} dp_{\rm cl}=1$. Differentiating $n$ times  (\ref{IntWeights}) with respect to $\eta$ yields
\begin{equation}
p_n=\frac{\frac{\hbar\omega_0}{k T_{\rm eff}}}{\left(1+\frac{\hbar\omega_0}{k T_{\rm eff}}\right)^n}
\,\propto\, e^{-n \ln\left(1+\frac{\hbar \omega_0}{k T_{\rm eff}}\right)}\,.
\label{eq:basci_power_law_prob}
\end{equation}
Following relation (\ref{QNRKB}) and the definition of 
$T_{\rm eff}$, one finds 
\begin{equation}
kT_{\rm eff}= \frac{\hbar \omega_0}{2}\, \bigg[\coth \bigg(\frac{\hbar \omega_0}{2k T_{\rm bath}}\bigg)-1\bigg]
\end{equation}
and
\begin{equation}
1+\frac{\hbar \omega_0}{k T_{\rm eff}}=
1+\frac{1}{\frac{\exp\left(-\frac{\hbar \omega_0}{2k T_{\rm bath}}\right)}{\exp\left(\frac{\hbar \omega_0}{2k T_{\rm bath}}\right)-
\exp\left(-\frac{\hbar \omega_0}{2k T_{\rm bath}}\right)}}=
\exp\left(-\frac{\hbar \omega_0}{k T_{\rm bath}}\right)\,.
\end{equation}
The ensuing expression for $p_n$ is 
\begin{equation}
p_n\propto e^{-\frac{n \hbar \omega_0}{k T_{\rm bath}}}
\propto e^{-\frac{E_n}{k T_{\rm bath}}}\,,
\end{equation}
which demonstrates that the distribution of the state weights is Boltzmannian if one uses a white noise with autocorrelation (\ref{QNRKB}), with, moreover, $T_{\rm sub}=T_{\rm bath}$ for
all coupling strengths and all bath temperatures. This reasoning can be easily extended to the 3D case.

\subsubsection{With other initial states}\label{HarmoTeff}

\begin{figure}[H]
 \centering
\begin{minipage}[b]{0.45\linewidth}
\iffigsdirectory
\includegraphics[height=50mm]{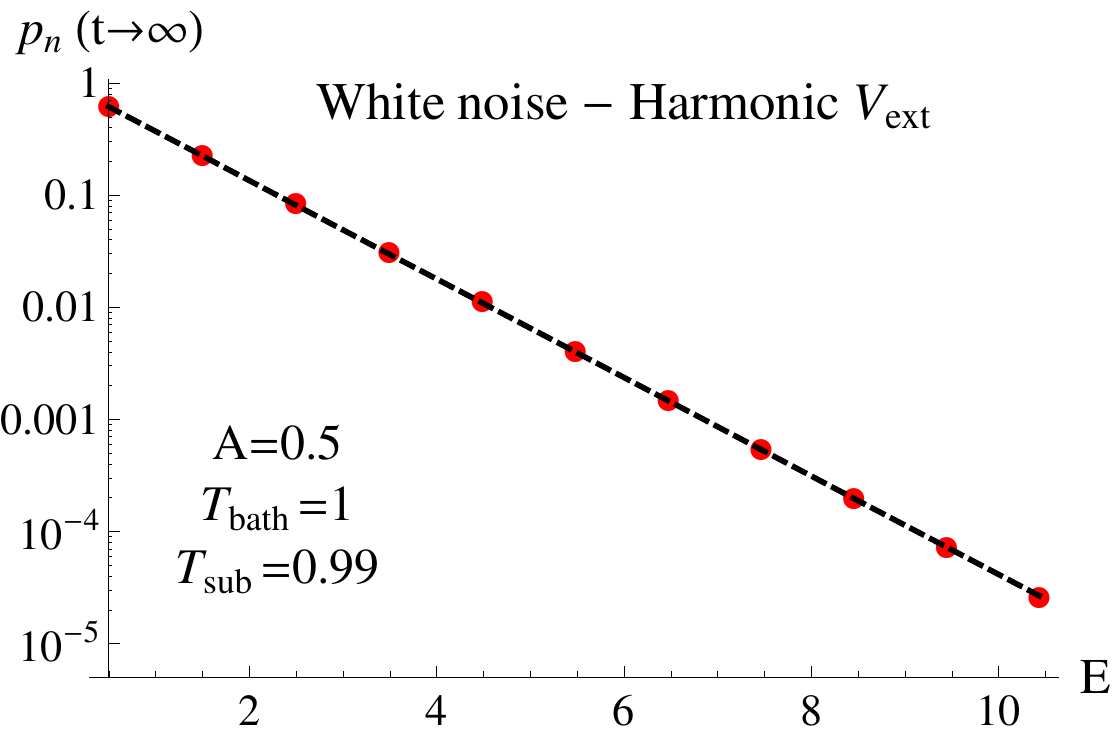}
\else
\includegraphics[height=50mm]{Psi0HarmoWeightsAsymp05.pdf}
\fi

 \caption{\label{fig: AsymHarmPn}
   \small  The asymptotic distribution of the eigenstate weights $p_{n=0,...10}$ (red dots), obtained with $A=0.5$ and  $T_{\rm bath}=1$, function of the corresponding eigenenergies $E_{n=0,...10}$. It fits the Boltzmann distribution ($\propto e^{-E/T_{\rm sub}}$) with $T_{\rm sub}=0.99$ (dashed line).}
\end{minipage}
\quad\,\,\,\,
\begin{minipage}[b]{0.45\linewidth}
\iffigsdirectory
\includegraphics[height=50mm]{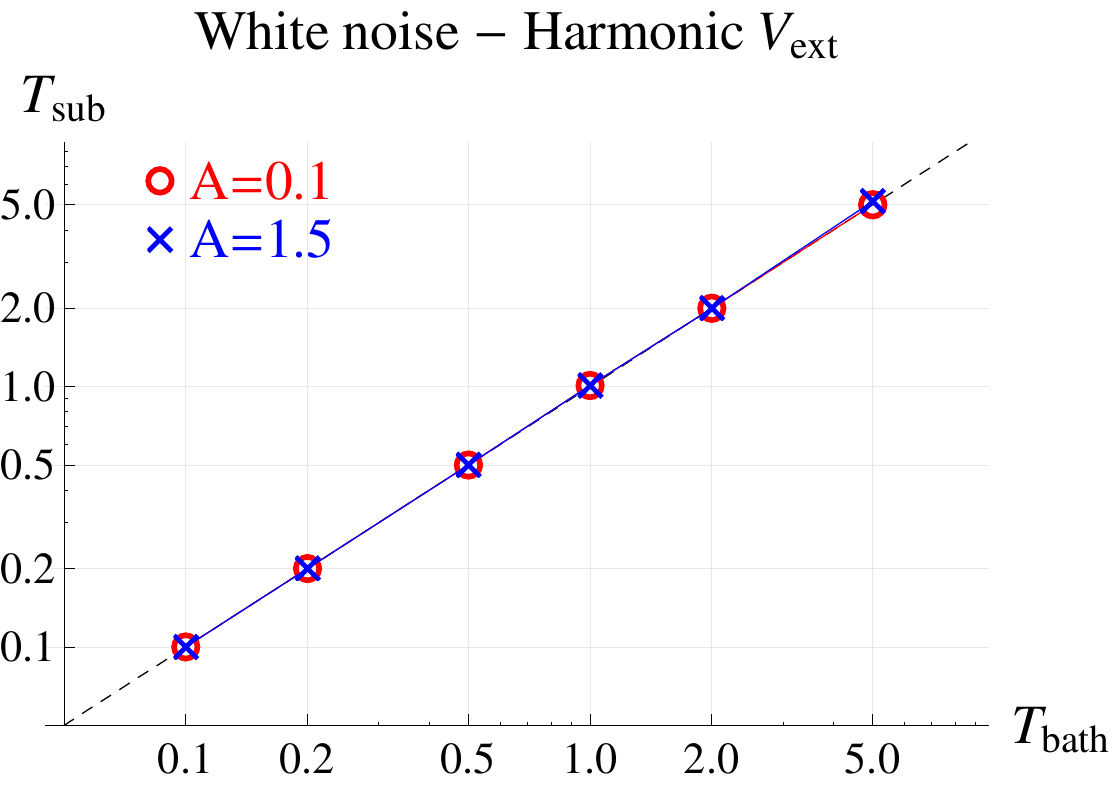}
\else
\includegraphics[height=50mm]{WhiteHarmoTeffvsT.pdf}
\fi
   \caption{\label{fig: HWTeffvsT}
   \small  Asymptotic subsystem temperature $T_{\rm sub}$ as a function of the bath temperature $T_{\rm bath}$ for two different friction coefficients: $A=0.1$ (red circles) and $A=1.5$ (blue crosses) corresponding respectively to a weak and strong coupling. The dashed line corresponds to the ideal case $T_{\rm sub}=T_{\rm bath}$.}
\end{minipage}
\end{figure}

We now perform the numerical simulations with the dimensionless SLE to generalize the previous results to other initial states. As shown for instance in Fig.~\ref{fig: AsymHarmPn}, the asymptotic distribution of the weights is independent of the chosen initial state and perfectly fits a Boltzmann distribution. One can determine the temperature effectively reached by the subsystem, $T_{\rm sub}$, by fitting the Boltzmann distribution $\propto e^{-E/T_{\rm sub}}$ to the asymptotic  $\{p_n(E_n)\}_{n=0,...10}$ values. For this particular example, one finds that $T_{\rm sub}=0.99 \simeq T_{\rm bath}$.

In Fig.~\ref{fig: HWTeffvsT}, we compare the temperature reached by our subsystem $T_{\rm sub}$ to the bath temperature  $T_{\rm bath}$ (defined as the temperature entering the noise correlation). For a large range of temperatures and independently of the friction coefficient $A$ and initial state, we observe that $T_{\rm sub}\simeq T_{\rm bath}$ and that the asymptotic distributions of the weights are Boltzmannian. One can thus conclude that the subsystem correctly thermalizes when one uses the white noise autocorrelation (\ref{QNRKB}). This is consistent 
with the assumption made in section \ref{Scenario} as regards the
asymptotic behavior of any initial state and the stochastic evolution of its centroids. In this respect, Figs.
\ref{fig: AsymHarmPn} and ~\ref{fig: HWTeffvsT} can also be understood
as sanity tests of our numerical procedures.

The total uncertainty on the asymptotic values, for a statistic of a few thousands of realizations, grows with the temperature from $\sim 2\%$ at  $T_{\rm bath}=0.1$ to $\sim 10\%$ at $T_{\rm bath}=5$. An additional averaging over a time range $\Delta t'$ once the equilibrium reached, leads to more reliable results whose accuracy then follows the common statistical law $\propto 1/\sqrt{n_{\rm stat}\times\Delta t'}$. 

To conclude, we have generalized the analytic results obtained in Sec.~\ref{AnalHarmoAsympt} with an initial Gaussian wave packet to other initial states. 
We can thus conjecture that within the case of the harmonic potential and the white noise autocorrelation (\ref{QNRKB}), the SLE universally leads to the thermal equilibrium of statistical mechanics. By universal, we mean that this result is independent of the bath temperature, friction coefficient and the initial state. Although this result was only expected at the weak coupling limit (as explained in the introduction), it is also reached in the intermediate and strong coupling regimes.

\subsection{Equilibration with a colored noise}
\label{Section3HarmoCol}

\begin{figure}[H]
 \centerline{
 \iffigsdirectory
 \includegraphics[height=50mm]{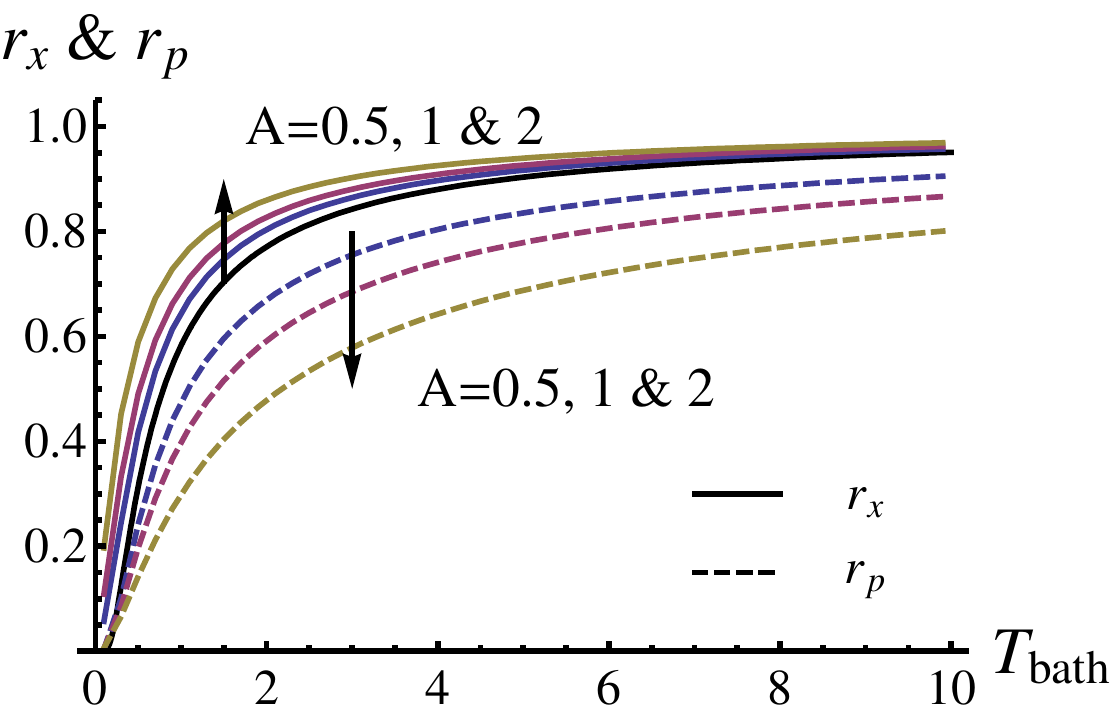}
 \else
 \includegraphics[height=50mm]{rx_and_rp.pdf}
 \fi
}
 \caption{
   \small The $r_x$ and $r_p$ reduction factors for several 
   friction coefficients $A$ (in units of $\omega_0$), as a function of $T_{\rm bath}$ (in units of $\hbar \omega_0/k$). One has generically $r_x(A',T)>r_x(A,T)$
   and $r_p(A',T)<r_p(A,T)$ for $A'>A$, as represented by the
   arrows.}
   \label{fig:rxandrpCol}
\end{figure}

We now use the colored noise (\ref{QNFordNormProd}) to study the SLE ability to bring a subsystem to the thermal equilibrium of statistical mechanics. We concentrate our study on the asymptotic distributions of weights, as other properties are similar to the ones discussed in the previous section devoted to the white noise case. For initial Gaussian wave packets, the equations governing the evolution of $\alpha$, $x_{\rm cl}$ and $p_{\rm cl}$ established in section \ref{AnalHarmo} hold as well, so that each realization of the evolution with a stochastic force 
leads to an asymptotic wave packet of width $\sqrt{\hbar/m\omega_0}$. It is thus possible to extend the method used in section \ref{AnalHarmoAsympt}.
For this purpose, one needs to evaluate the asymptotic distribution $W(x_{\rm cl},p_{\rm cl})$ for the case of a classical damped harmonic oscillator driven by some colored noise. This can be done in several steps. One first decouples the equations of motion (\ref{xpPFD}) by setting
\begin{equation}
\left(\begin{array}{c}
x_{\rm cl}(t) \\ p_{\rm cl}(t)
\end{array}\right)  = c_+(t)  v_+ + c_-(t) v_-  \,,
\end{equation}
where $v_+$ and $v_-$ are eigenvectors of the matrix 
$\left(\begin{array}{cc}
0 & -1/m \\ m \omega_0^2 & A
\end{array}\right)$ with respective eigenvalues
$\lambda_\pm = \frac{A}{2}\pm \sqrt{\frac{A^2}{4}-\omega_0^2}$: 
$v_\pm=\left(\begin{array}{c}1\\ -m \lambda_{\pm}
\end{array}\right)$. The equations of motion for $c_{\pm}$ then write
\begin{eqnarray}
\dot{c}_{+}=-\lambda_+ c_+ - \frac{F_R}{m\sqrt{A^2-4 \omega_0^2}}
\quad\text{and}\quad \dot{c}_{-}=-\lambda_- c_- + \frac{F_R}{m\sqrt{A^2-4 \omega_0^2}}\,.
\end{eqnarray} 
Each of these equations describes the motion of a free particle
undergoing ohmic friction and colored noise and admits a trivial solution, e.g.
\begin{equation}
c_+(t)= - \frac{1}{m\sqrt{A^2-4 \omega_0^2}}
\int_{0}^{t} e^{-\lambda_+(t-t')} F_R(t')\,.
\label{eq_cplus_evol}
\end{equation}
In the asymptotic limit, the autocorrelations between $x_{\rm cl}$ and $p_{\rm cl}$ can thus be generated from those between $c_+$ and $c_-$.
In particular, one obtains
\begin{equation}
\left(\begin{array}{cc}
\langle x_{\rm cl}^2 \rangle & \langle x_{\rm cl} p_{\rm cl} \rangle
\\
\langle x_{\rm cl} p_{\rm cl} \rangle & \langle p_{\rm cl}^2 \rangle
\end{array}\right) = 
\langle c_+^2 \rangle v_+^T \otimes v_+ +
\langle c_+ c_- \rangle (v_+^T \otimes v_- +  
v_-^T \otimes v_+) +
\langle c_-^2 \rangle v_-^T \otimes v_-\,. 
\end{equation}
Using equation (\ref{eq_cplus_evol}) as well as the corresponding one
for $c_-$ evolution, one obtains
\begin{equation}
\langle c_\pm^2\rangle_{\rm asympt}
=\frac{1}{m^2(A^2-4\omega_0^2)}\int_0^{+\infty} dt_1
\int_0^{+\infty} dt_2\, e^{-\lambda_\pm (t_1+t_2)} C(t_1-t_2)\,,
\end{equation}
where $C(\tau)$ is the force autocorrelation $\langle F_R(t) F_R(t+\tau)\rangle$. Proceeding to the variable changes $\Sigma=t_1+t_2$ and $\tau=|t_1-t_2|$ leads to
\begin{equation}
\langle c_\pm^2\rangle_{\rm asympt}
= \frac{\mathcal{L}_C(\lambda_\pm)}{m^2(A^2-4\omega_0^2)\lambda_\pm}\,,
\end{equation}
where $\mathcal{L}_C$ designates the Laplace transform of $C$, while 
similar calculation for $\langle x_{\rm cl} p_{\rm cl}\rangle$ gives 
\begin{equation}
\langle x_{\rm cl} p_{\rm cl}\rangle_{\rm asympt}=
-\frac{\mathcal{L}_C(\lambda_+)+\mathcal{L}_C(\lambda_-)}
{m^2 A(A^2-4\omega_0^2)}\,.
\end{equation}
Gathering all results, the asymptotic correlations write  
\begin{eqnarray}
\langle x_{\rm cl}^2\rangle_{\rm asympt}&=&
\frac{1}{A m^2 \omega_0^2 \sqrt{A^2-4 \omega_0^2}}
\left(\lambda_+  \mathcal{L}_C(\lambda_-) - \lambda_-
 \mathcal{L}_C(\lambda_+) \right)
 \nonumber\\
\langle p_{\rm cl}^2\rangle_{\rm asympt}&=&
\frac{1}{A \sqrt{A^2-4 \omega_0^2}}
\left(\lambda_+  \mathcal{L}_C(\lambda_+) - \lambda_-
 \mathcal{L}_C(\lambda_-) \right) 
  \nonumber\\
 \langle x_{\rm cl}  p_{\rm cl}\rangle_{\rm asympt}&=&0\,.
\end{eqnarray}
For $A<2 \omega_0$ one obtains complex eigenvalues with however 
$\lambda_-=\bar{\lambda}_+$ and $\mathcal{L}_C(\lambda_-)=
\bar{\mathcal{L}}_C(\lambda_+)$, which guarantees that both $\langle x_{\rm cl}^2\rangle_{\rm asympt}$ and $\langle p_{\rm cl}^2\rangle_{\rm asympt}$ are genuine real quantities:
\begin{eqnarray}
\langle x_{\rm cl}^2\rangle_{\rm asympt}&=&
\frac{1}{m^2 \omega_0^2}\left(
\frac{\Re  \mathcal{L}_C(\lambda_+)}{A}-
\frac{\Im  \mathcal{L}_C(\lambda_+)}{\sqrt{4 \omega_0^2-A^2}}
\right)\nonumber\\
\langle p_{\rm cl}^2\rangle_{\rm asympt}&=&
\frac{\Re  \mathcal{L}_C(\lambda_+)}{A}+
\frac{\Im  \mathcal{L}_C(\lambda_+)}{\sqrt{4 \omega_0^2-A^2}}\,.
\label{eq:x2_and_p2_from_real_and_imag}
\end{eqnarray}
In the white noise case limit $\sigma \rightarrow 0$, one has $C(\tau)=B \delta(\tau)$ with 
$B= 2 m A k T_{\rm bath}$ according to Einstein relation\footnote{
One easily checks that this is true for the colored noise (\ref{QNFordNormProd}) when $T_{\rm bath} \rightarrow +\infty$.}. As a consequence, 
$\mathcal{L}_C(\lambda_\pm)= m A k T_{\rm bath}$ and one recovers 
$\langle x_{\rm cl}^2\rangle_{\rm asympt}=\frac{kT_{\rm bath}}{m \omega_0^2}$
as well as 
$\langle p_{\rm cl}^2\rangle_{\rm asympt}= m kT_{\rm bath}$, so that each 
degree of freedom carries the expected average energy 
$\frac{kT_{\rm bath}}{2}$. For the general case we express
\begin{equation}
\langle x_{\rm cl}^2\rangle_{\rm asympt} =
\frac{kT_{\rm bath}}{m \omega_0^2} \times r_x \quad\text{and}\quad
\langle p_{\rm cl}^2\rangle_{\rm asympt} =
m kT_{\rm bath} \times r_p\,,
\end{equation}
where
\begin{equation}
r_x =
\frac{\lambda_+  \mathcal{L}_C(\lambda_-) - \lambda_-
 \mathcal{L}_C(\lambda_+) }{m  A kT_{\rm bath}\sqrt{A^2-4 \omega_0^2}}
\quad\text{and}\quad
r_p =
\frac{\lambda_+  \mathcal{L}_C(\lambda_+) - \lambda_-
 \mathcal{L}_C(\lambda_-)}{m A kT_{\rm bath} \sqrt{A^2-4 \omega_0^2}}\,.
\end{equation}
are reduction factors which encodes the deviations with respect to the white noise case. Finally, one easily demonstrates that only the second cumulants
$\langle x_{\rm cl}^2\rangle_{\rm asympt}$ and  
$\langle x_{\rm cl}^2\rangle_{\rm asympt}$ are non-zero for stochastic forces $F_R$ of Gaussian nature. The ensuing asymptotic distribution of the centroids $(x_{\rm cl},p_{\rm cl})$ thus admits the simple form
\begin{equation}
W(x_{\rm cl},p_{\rm cl}) \propto e^{-\frac{m\omega_0^2x_{\rm cl}^2}{2  k T_{\rm bath} r_x}-\frac{p_{\rm cl}^2}{2m  k T_{\rm bath}r_p}}\,.
\end{equation}
For the noise autocorrelation (\ref{QNFordNormProd}),
one has 
\begin{equation}
\mathcal{L}_C(\lambda)=
\frac{m A}{\pi} \int_0^\infty \frac{\hbar \omega}{\exp(\hbar \omega/kT_{\rm bath})-1} \left(\frac{i}{\omega + i \lambda}
-\frac{i}{\omega - i \lambda}\right)\, d\omega
\label{eq:lapl_ford}
\end{equation}
so that $r_x$ and $r_p$ only depend on the dimensionless 
quantities $k T_{\rm bath}/\hbar \omega_0$ and $A/\omega_0$.
To our knowledge, the integral (\ref{eq:lapl_ford}) has no simple expression. It however admits two interesting limiting cases:
\begin{itemize}
\item High temperature: When $k T_{\rm bath} \gg \hbar |\lambda_{\pm}|$, $\exp(\hbar \omega/kT_{\rm bath})$ can be approximated by  $1 + \hbar \omega/kT_{\rm bath}$ in the integral, leading to $\mathcal{L}_C(\lambda)\approx
m A k T_{\rm bath}$ and $r_x\approx r_p\approx 1$: one 
reaches the classical limit.
\item Brownian hierarchy: When such hierarchy is satisfied $A^{-1}$ should be larger than any time scale of the subsystem. Under such conditions, $A\ll \omega_0$, $\lambda_+ \approx i \omega_0 + \frac{A}{2}$ and 
\begin{equation}
\mathcal{L}_C(\lambda_+)\approx
\frac{m A}{\pi} \int_0^\infty \frac{\hbar \omega}{\exp(\hbar \omega/kT_{\rm bath})-1} \left(\frac{i}{\omega -\omega_0 + i A/2}
-\frac{i}{\omega +\omega_0 - i A/2}\right)\, d\omega\,.
\end{equation} 
One has $A \ll k T_{\rm bath}/\hbar$ as well and the poles $\pm(\omega_0-
i A/2)$ can then be considered to be infinitely close to the real axis, with $\frac{i}{\omega -\omega_0 + i A/2}=i {\rm p.v.}
\left(\frac{1}{\omega-\omega_0}\right)+ \pi \delta(\omega-\omega_0)$
and   $\frac{i}{\omega +\omega_0 - i A/2}=\frac{i}{\omega+ \omega_0}$,
resulting in 
\begin{equation}
\mathcal{L}_C(\lambda_+)\approx
\frac{m A \hbar \omega_0}{\exp(\hbar \omega_0/kT_{\rm bath})-1}
+ \frac{i m A}{\pi} {\rm p.v.}\int_0^\infty \frac{\hbar \omega}{\exp(\hbar \omega/kT_{\rm bath})-1}\,\frac{2 \omega_0}{\omega^2 -\omega_0^2}\, d\omega\,.
\end{equation}
According to  equation (\ref{eq:x2_and_p2_from_real_and_imag}), the small $A$ factor showing up in the real part of 
$\mathcal{L}_C(\lambda_+)$ is compensated by an equivalent factor
in the denominator. Therefore   
\begin{equation}
\langle x_{\rm cl}^2\rangle_{\rm asympt}=
\frac{\langle p_{\rm cl}^2\rangle_{\rm asympt}}{m^2 \omega_0^2}
= \frac{1}{m \omega_0^2}\,
\frac{\hbar \omega_0}{\exp(\hbar \omega_0/kT_{\rm bath})-1}
\end{equation}
and 
\begin{equation}
r_x=r_p=\frac{\hbar \omega_0/kT_{\rm bath}}
{\exp(\hbar \omega_0/kT_{\rm bath})-1}\,.
\label{eq:rx_and_rp}
\end{equation}
\end{itemize}
On Fig.~\ref{fig:rxandrpCol} we illustrate the quantities $r_x$ and $r_p$ as a function of $k T_{\rm bath}/\hbar \omega_0$ for several values of $A/\omega_0$. One notices that $r_p < r_x<1$ for finite $A/\omega_0$, which breaks the equipartition of energy.

Following the method of section \ref{AnalHarmoAsympt}, the asymptotic
weight of the $n^{\rm th}$ state writes
\begin{equation}
p_n =\frac{(-1)^n}{n!} \left.\frac{\partial^n}{\partial \eta^n}
G(\omega_0,A,T_{\rm bath})\right|_{\eta=1}\,,
\label{eq:pn_def_colored}
\end{equation}
where the generating function $G$ is defined as
\begin{equation}
G(\omega_0,A,T_{\rm bath})=
\int 
W(x_{\rm cl},p_{\rm cl}) e^{-\left(\frac{(x_{\rm cl}/a)^2}{2}+
\frac{(p_{\rm cl} a/\hbar)^2}{2}\right)}
d x_{\rm cl} d p_{\rm cl} 
 =\frac{\Pi_x^{\frac{1}{2}}(0)\Pi_p^{\frac{1}{2}}(0)}{\Pi_x^{\frac{1}{2}}(\eta)\Pi_p^{\frac{1}{2}}(\eta)}\,,
\end{equation}
with 
\begin{equation}
\Pi_{x}(\eta)= \eta + \frac{\hbar \omega_0}{k T_{\rm bath} r_x}
\quad\text{and}\quad
\Pi_{p}(\eta)= \eta + \frac{\hbar \omega_0}{k T_{\rm bath} r_p}\,.
\end{equation}
Quite generally, $r_x\neq r_p$ and the $p_n$ generated from $G$ are 
more involved than the simple power law  $p_n \propto c^{-n}$ found in equation (\ref{eq:basci_power_law_prob}) for the white noise case. As a consequence, deviations from usual Boltzmann distributions are expected for the $p_n$. Noticeable exceptions are:
\begin{itemize} 
\item
the case of a classical noise obtained for large $T_{\rm bath}$, for which
$r_x=r_p=1$ and $p_n\propto \exp\left(-n \ln\left(1+\frac{\hbar \omega _0}{k T_{\rm bath}}\right)\right)\approx \exp\left(-E_n/k T_{\rm bath}\right)$
\item the weak coupling case obtained at small $A$ for which,
according to equation (\ref{eq:rx_and_rp}),  
$\Pi_x=\Pi_p=\eta-1+\exp(\hbar \omega_0 / k T_{\rm bath})$ implying 
that $p_{n} \propto \exp(- n \hbar \omega_0 / k T_{\rm bath}) 
\propto \exp(-E_n/ k T_{\rm bath})$ as well.
\end{itemize}
In these two cases, the temperature governing the distribution of weights for the subsystem ($T_{\rm sub}$) is found to be exactly
the bath temperature $T_{\rm bath}$. 

\begin{figure}[H]
 \centerline{
 \iffigsdirectory
\includegraphics[height=45mm]{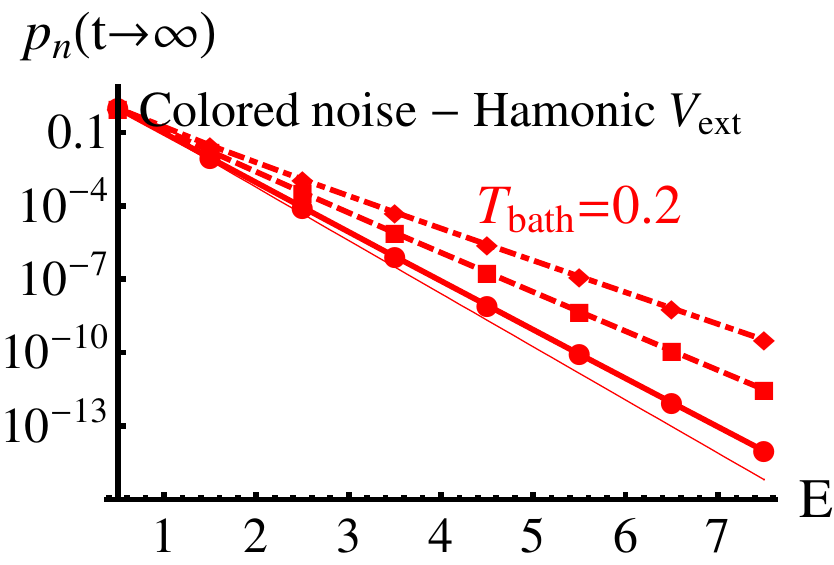}
\includegraphics[height=45mm]{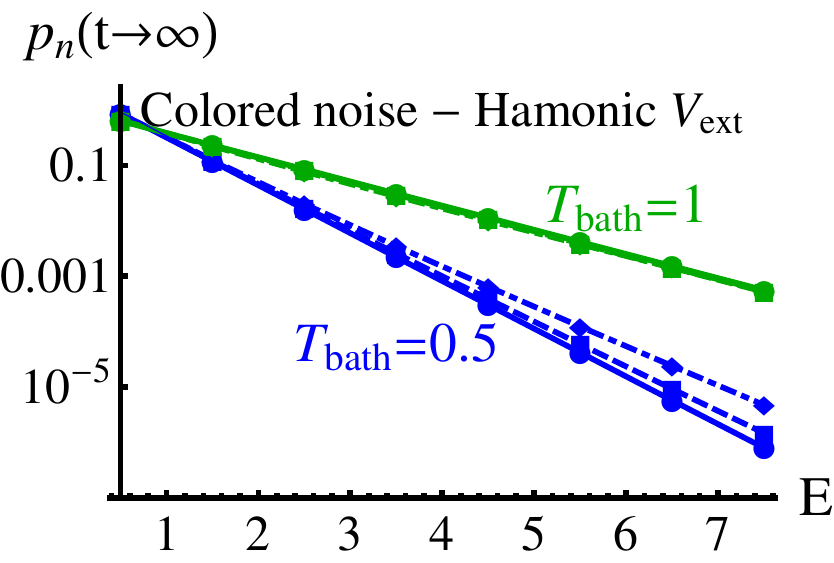} 
 \else
\includegraphics[height=45mm]{HarmoWeightsAsymptT02Colored.pdf}
\includegraphics[height=45mm]{HarmoWeightsAsymptT05andT1Colored.pdf} 
 \fi
}
   \caption{
   \small  The asymptotic distributions of the eigenstate weights $p_{n=0,...8}$ (joined by lines) function of the eigenenergies $E_{n=0,...8}$, obtained with different friction coefficients, measured in units of $\omega_0$: $A=0.1$ (solid lines), $A=0.5$ (dashed lines) and $A=1.5$ (dot-dashed lines) and temperatures (in units of $\hbar \omega_0/k$): $T_{\rm bath}=0.2$ ({\it left}), $T_{\rm bath}=0.5$ and $1$ ({\it right}). They are compared to the corresponding ``ideal" Boltzmann distributions $\propto e^{-E/T_{\rm bath}}$ (thin lines). }
   \label{fig:AsymHarmoPnCol}
\end{figure} 

Those various aspects are well illustrated on 
Fig.~\ref{fig:AsymHarmoPnCol}, where deviations with respect to the
Boltzmann distribution $p_n\propto \exp(-E_n/T_{\rm bath})$
are observed for "large" friction coefficients or "small" temperatures.
For these cases, the distribution of weights appears to be better described
by the law $p_n\propto \exp(-E_n/T_{\rm sub})$, i.e. by introducing 
some effective temperature $T_{\rm sub}$ specific to the subsystem, as explained in the introduction. Yet, the definition of $T_{\rm sub}$
is not unique as the $\{p_n\}$ shows genuine deviations from a power
law. As one is often interested in the low lying eigenstates in phenomenology (the fundamental and few lower excited eigenstates), a {\it bona fide} choice for $T_{\rm sub}$ will be adopted here as
\begin{equation}
T_{\rm sub}\left(\{p_n\}\right)= -\frac{E_1-E_0}{\ln(p_1/p_0)}\,.
\label{def:Tsub}
\end{equation}
For the harmonic potential, $p_0$ and $p_1$ are evaluated thanks to equation (\ref{eq:pn_def_colored}) and definition 
(\ref{def:Tsub}) leads to 
\begin{equation}
T_{\rm sub}= -\frac{1}{\ln \left(\frac{1}{2}\left(\frac{1}{1+\frac{1}{T_{\rm bath} r_x}}+\frac{1}{1+\frac{1}{T_{\rm bath} r_p}}\right)\right)}\,.
\end{equation}

\begin{figure}[H]
\centering
\iffigsdirectory
\includegraphics[height=50mm]{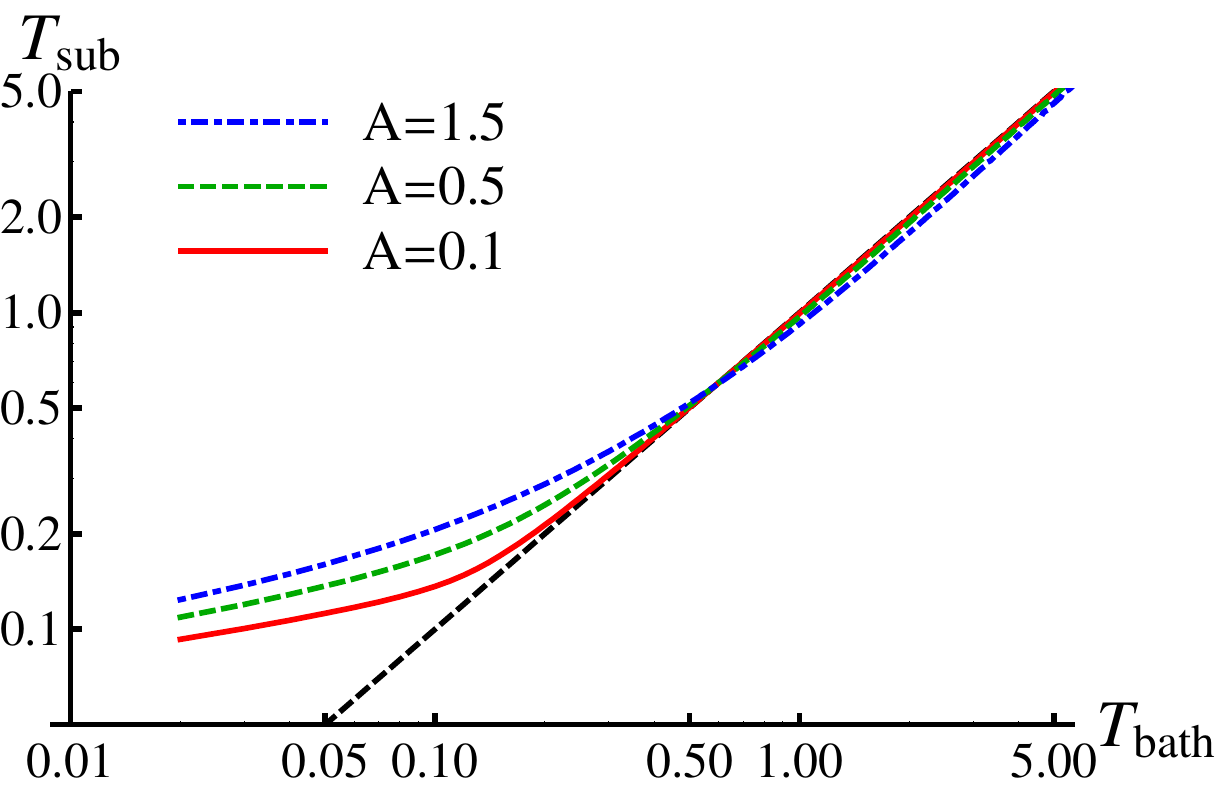}
\else
\includegraphics[height=50mm]{ColoredHarmoTeffvsT2.pdf}
\fi
\caption{
   \small  Asymptotic subsystem temperature $T_{\rm sub}$ as a function of the bath temperature $T_{\rm bath}$ -- both in units of $\hbar \omega_0/k$ -- for three different friction coefficients (in units of $\omega_0$): $A=0.1$ (solid line), $A=0.5$ (dashed line) and $A=1.5$ (dot-dashed line) corresponding respectively to a weak, intermediate and strong coupling. The straight dashed line corresponds to the ideal case $T_{\rm sub}=T_{\rm bath}$. }
\label{fig: HColTeffvsT}
\end{figure} 

In Fig.~\ref{fig: HColTeffvsT}, we compare this effective temperature actually reached by our subsystem $T_{\rm sub}$ to the bath temperature $T_{\rm bath}$ used as input of the noise. When $T_{\rm bath}\gtrsim 0.5$, the subsystem thermalizes with $T_{\rm sub}\approx T_{\rm bath}$ in a good approximation (we note a slight dependence vs the friction coefficient: the larger $A$ the smaller $T_{\rm sub}$). At lower temperatures, some discrepancies appear in the form of a saturation of $T_{\rm sub}$. If $A\lesssim \omega_0$ the onset of those deviations is however delayed until $T_{\rm bath} \lesssim A$ where the Brownian hierarchy is then broken.

Using our numerical tool, we have also studied the case where the 
initial state is not a Gaussian wave packet. Our conclusion is that 
the asymptotic states always turn out to be Gaussian wave packets of
width $\sqrt{\hbar/m\omega_0}$ (which corresponds to the ground state 
width). We thus conjecture that the main results obtained in this section for the harmonic potential are independent of the initial state. 

\section{Equilibration with a linear oscillator}
\label{Section4Linear}

In this section, we study the thermal relaxation given by the SLE (\ref{SLeq}) with the linear 1D potential $V_{\rm ext}=K_l\,|x|/2$ and with the white (\ref{QNRK}) or colored (\ref{QNFordNormProd}) noise. The linear oscillator allows us to test the SLE for a non-harmonic situation. The analytic resolution being far from obvious in this case, we perform the analyses resorting to numerical simulations. 
For all initial states investigated, we have found similar asymptotic features and we will therefore not discuss the role of the initial state any further.

\subsection{Equilibration with a white noise}
\label{Section4LinearWhite}

As explained in Sec.~\ref{QNchoice}, the white noise (\ref{QNRKB}) was initially derived for a harmonic potential. In this section, we test its ability to be extended to other types of potentials through the example of the linear potential. In the white noise expression (\ref{QNRKB}),
we set $E_0$ to the corresponding ground state energy, i.e. $E_0\approx 0.509\,\hbar\omega_0$, with $\hbar\omega_0=(K_l \hbar)^{2/3}/m^{1/3}$.

\begin{figure}[!h]\label{fig: AvEnLinW}
 \centerline{
 \iffigsdirectory
 \includegraphics[height=50mm]{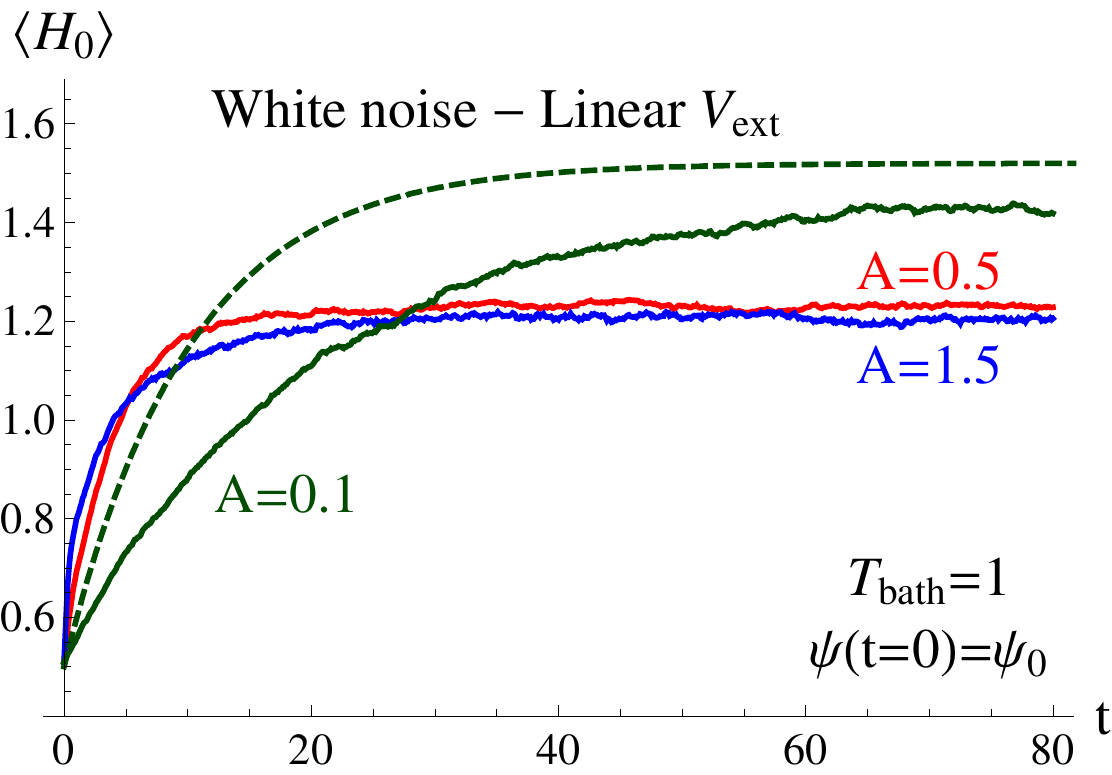}
 \else
 \includegraphics[height=50mm]{AverageEnergyEvolLinWhite.pdf}
 \fi}
   \caption{\label{fig: AvEnLinW}
   \small  Numerical average energy $\langle H_0\rangle$ evolutions for different friction coefficients $A$ (solid curves) and the theoretical evolution given by (\ref{SenEnerEvol}) with $ T_{\rm bath}=1$, $\langle H_0\rangle(t\rightarrow\infty)=1.52$ and $A=0.1$ (dashed curve). }
\end{figure} 

We first consider the average energy observable  $\langle H_0 \rangle$. As shown in Fig.~\ref{fig: AvEnLinW}, the asymptotic value of $\langle H_0 \rangle$ differs
in general from the expected value 
\begin{eqnarray}\label{H0AsympLin}
\langle H_0\rangle(t\rightarrow\infty)=\frac{\sum_i E_i\,e^{-E_i/T_{\rm bath}}}{\sum_i e^{-E_i/T_{\rm bath}}}\simeq 1.52\,,
\end{eqnarray}
and exhibits a strong $A$-dependence, in contrast
with the harmonic potential case. For small friction coefficients ($A<1$), the average energy evolutions are nevertheless in good agreement with the exponential rate (\ref{SenEnerEvol}) when one takes the actual $\langle H_0\rangle(t\rightarrow\infty)$ and effective $A_{\rm eff}\simeq A/2$ values.\\

Independent of the initial state, the asymptotic distributions of the weights $p_{n=0,...10}$ are close to the Boltzmann distributions $\propto e^{-E/T_{\rm bath}}$ only when $1\lesssim T_{\rm bath}\lesssim 2$ at weak couplings (see Fig.~\ref{fig: AsymLinPn}). At low temperatures strong discrepancies are observed: the higher excited states exceed the Boltzmann law, exhibit an alternating pattern and saturate at low weights. 
 A dependence on the friction coefficient value is observed from the $2^{\rm nd}$ ($4^{\rm th}$) excited state at low (medium) temperatures. The latter explains the $\langle H_0\rangle$ dependence on the friction coefficient observed in Fig.~\ref{fig: AvEnLinW}: a smaller friction coefficient is observed to generate higher populations for the excited eigenstates and thus a higher average energy. At large temperatures $T_{\rm bath}\gtrsim 5$, the distribution of weights is difficult to evaluate because of statistical fluctuations and numerical scheme imperfections.

\begin{figure}[H]\label{fig: AsymLinPn}
 \centerline{
 \iffigsdirectory
 \includegraphics[height=50mm]{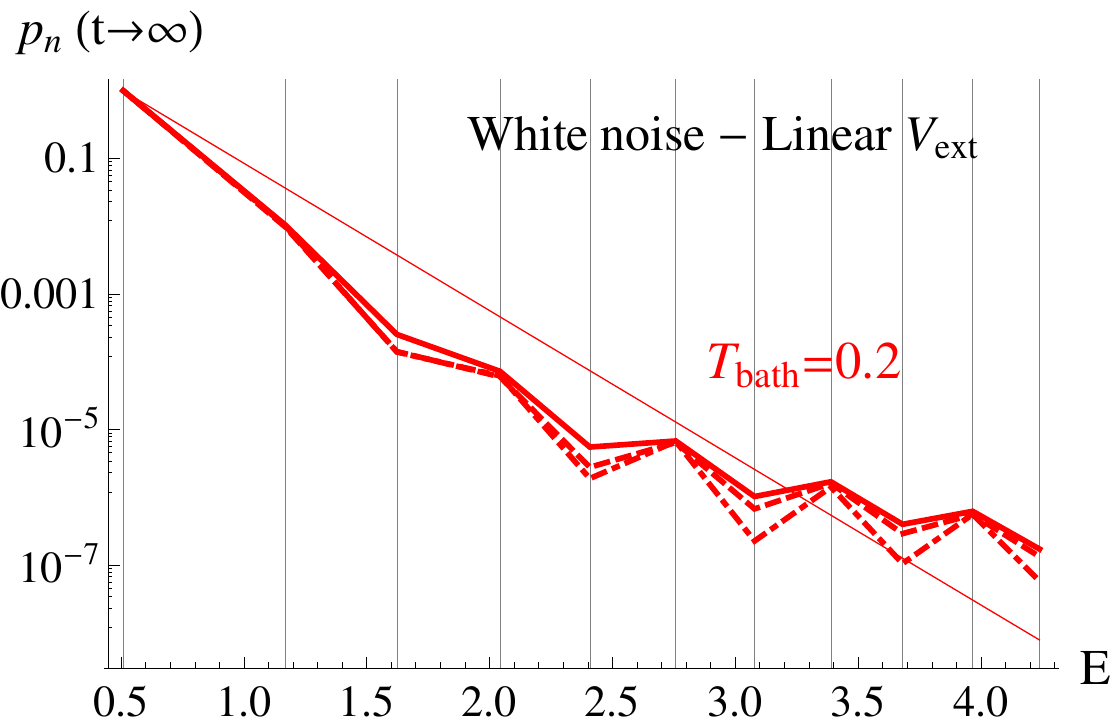}
 \includegraphics[height=50mm]{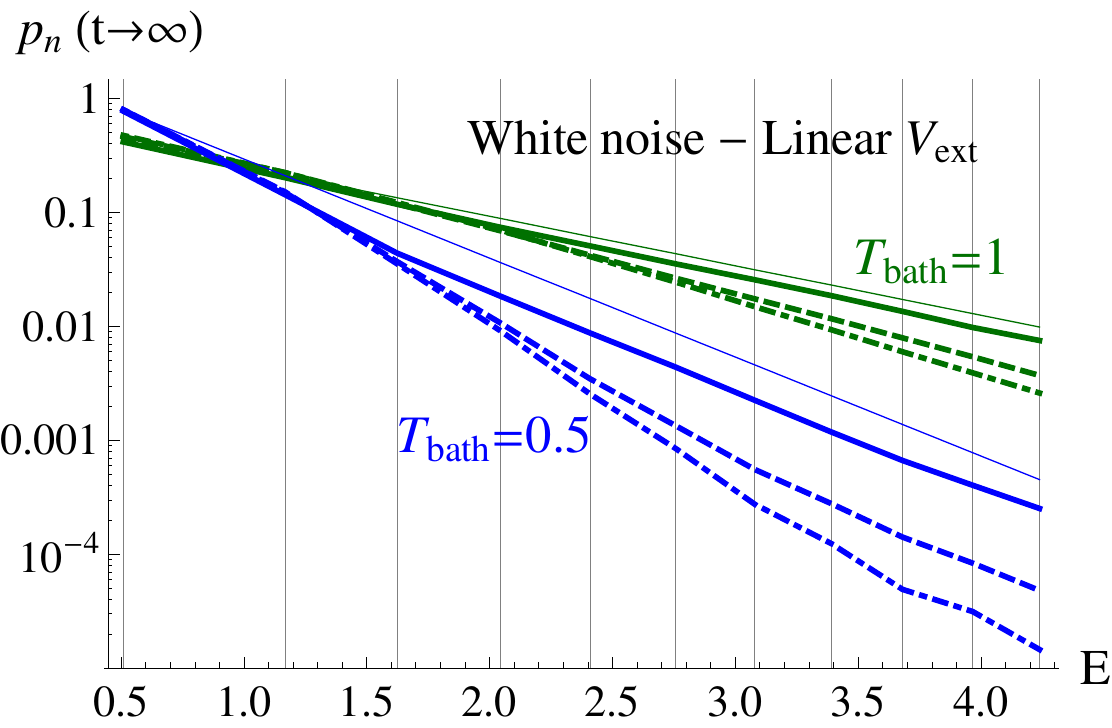}
 \else
 \includegraphics[height=50mm]{LinWeightsAsympT02.pdf}
 \includegraphics[height=50mm]{LinWeightsAsympT05and1.pdf}
 \fi}
   \caption{\label{fig: AsymLinPn}
   \small  The asymptotic distributions of the eigenstate weights $p_{n=0,...10}$ (joined by lines) function of the eigenenergies $E_{n=0,...10}$ (vertical lines), obtained with different friction coefficients $A=0.1$ (solid lines), $A=0.5$ (dashed lines) and $A=1.5$ (dot-dashed lines) and temperatures $T_{\rm bath}=0.2$ ({\it left}), $T_{\rm bath}=0.5$ and $1$ ({\it right}). They are compared to the corresponding ``ideal" Boltzmann distributions $\propto e^{-E/T_{\rm bath}}$ (thin lines). }
\end{figure} 
 
Given that the SLE does not lead to genuine Boltzmann distributions in this case, we will resort our definition (\ref{def:Tsub}) of the subsystem effective temperature $T_{\rm sub}$, which is obtained 
by fitting the Boltzmann law to the first two weights ($p_0$ and $p_1$) in lin-log space. The evaluation of such $T_{\rm sub}$ vs. $T_{\rm bath}$ is shown in Fig.~\ref{fig: TeffvsT}.
For any value of the friction coefficient, one observes clear deviations from the ``ideal" $T_{\rm sub}=T_{\rm bath}$ line at low and high temperatures, which could be interpreted as some
inefficiency of the SLE to heat up the subsystem. In particular, one does not recover this identity {\em even in the small coupling limit}. It appears in fact that the $T_{\rm sub}(T_{\rm bath})$ law is rather insensitive to $A$.

 \begin{figure}[!h]\label{fig: TeffvsT}
 \centerline{
 \iffigsdirectory
\includegraphics[height=50mm]{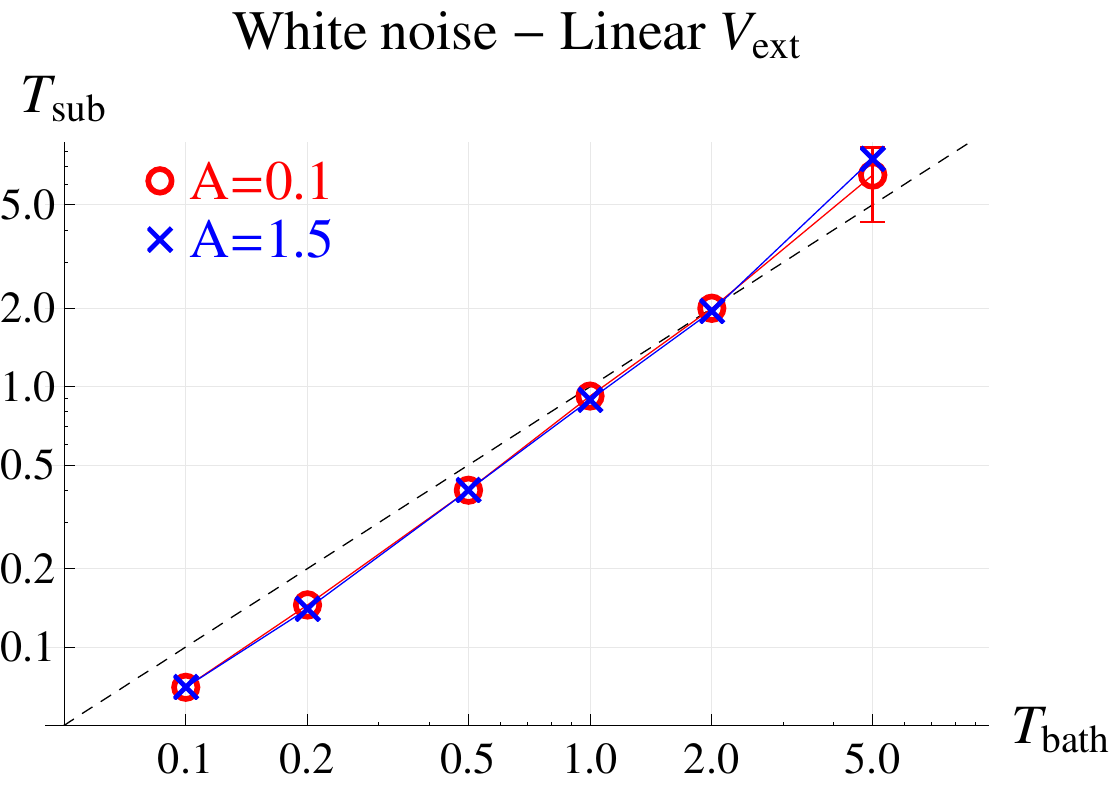}
 \else
\includegraphics[height=50mm]{WhiteLinearTeffvsT.pdf}
 \fi}
   \caption{\label{fig: TeffvsT}
   \small  Asymptotic subsystem temperature $T_{\rm sub}$ as a function of the bath temperature $T_{\rm bath}$ for two different friction coefficients $A=0.1$ (red circles) and $A=1.5$ (blue crosses) corresponding respectively to weak and strong couplings. The dashed line corresponds to the ideal case $T_{\rm sub}=T_{\rm bath}$.  
  At high temperatures our accuracy on $T_{\rm sub}$ is low due to a very large time required to reach the asymptotic behavior and a large uncertainty as in Sec.~\ref{HarmoTeff}.}
\end{figure} 
In Tab.~\ref{WhiteLinBoltz}, we assess 
the relevance of the effective Boltzmann laws $e^{-E_n/T_{\rm sub}}$ obtained once $T_{\rm sub}$ has been defined, by counting the numbers of low lying eigenstates which are close enough to this law. At low temperature, only a limited set of weights are found to be encompassed by this law.

\begin{table}[h!]
\begin{center}
    \begin{tabular}{|C{4.5cm}||C{2.5cm}|C{2.5cm}|C{2.5cm}|}
    \hline
\multicolumn{4}{ |C{11.8cm}| }{--- Number of weights close to $\propto e^{-E/T_{\rm sub}}$ ? ---}\\
    \hline 
$T_{\rm bath}$ $\backslash$  $A$ & Small & Intermediate & Large\\
    \hline
    \hline
Low ($T_{\rm bath}<0.5$) & 3 & 2 & 2 \\
    \hline
Medium ($0.5<T_{\rm bath}<2$) & 5 & 5 & 4  \\ 
    \hline
High ($T_{\rm bath}>2$) & 10 & 9 &  8 \\ 
    \hline
    \end{tabular}
\caption {\label{WhiteLinBoltz} 
\small Approximate number of weights close to the corresponding Boltzmannian $\propto e^{-E/T_{\rm sub}}$. One can consider the agreement to be poor from 2 to 4 weights, good from 5 to 7 and very good from 8 to 11. A better agreement is obtained toward the weak and/or high temperature regimes.}
\end{center}
\vspace{-0.4cm}
\end {table}

In view of these elements, we conclude that the white noise (\ref{QNRK}) is not quite suitable to obtain an acceptable thermal equilibrium (in the sense of $p_n\propto e^{-E_n/T_{\rm bath}}$) with other potentials than the harmonic one. Nevertheless, if one is interested in a limited number of low lying eigenstates (see Tab.~\ref{WhiteLinBoltz}), this formalism could be used for phenomenological purposes by performing a rescaling in the noise expression (\ref{QNRKB}): either by changing the value of $E_0$ (optimal value is found to be $E_0=0.33$) or by choosing the input $\tilde{T}_{\rm bath}$ such as to obtain the desired $T_{\rm sub}=T_{\rm bath}$. Conversely, this study confirms the very specific nature of the harmonic potential upon which general conclusions should not be drawn as regards the applicability of any scheme aiming at describing the thermalization of quantum subsystem.

\subsection{Equilibration with a colored noise}
\label{Section4LinearCol}

Unlike the white noise, the colored noise (\ref{QNFordNormProd}) was derived without assumptions on the potential. In this section, we test its ability to be extended to other potentials through the example of the linear potential. 
\begin{figure}[!h]\label{fig: AvEnLinCol}
 \centerline{
 \iffigsdirectory
 \includegraphics[height=50mm]{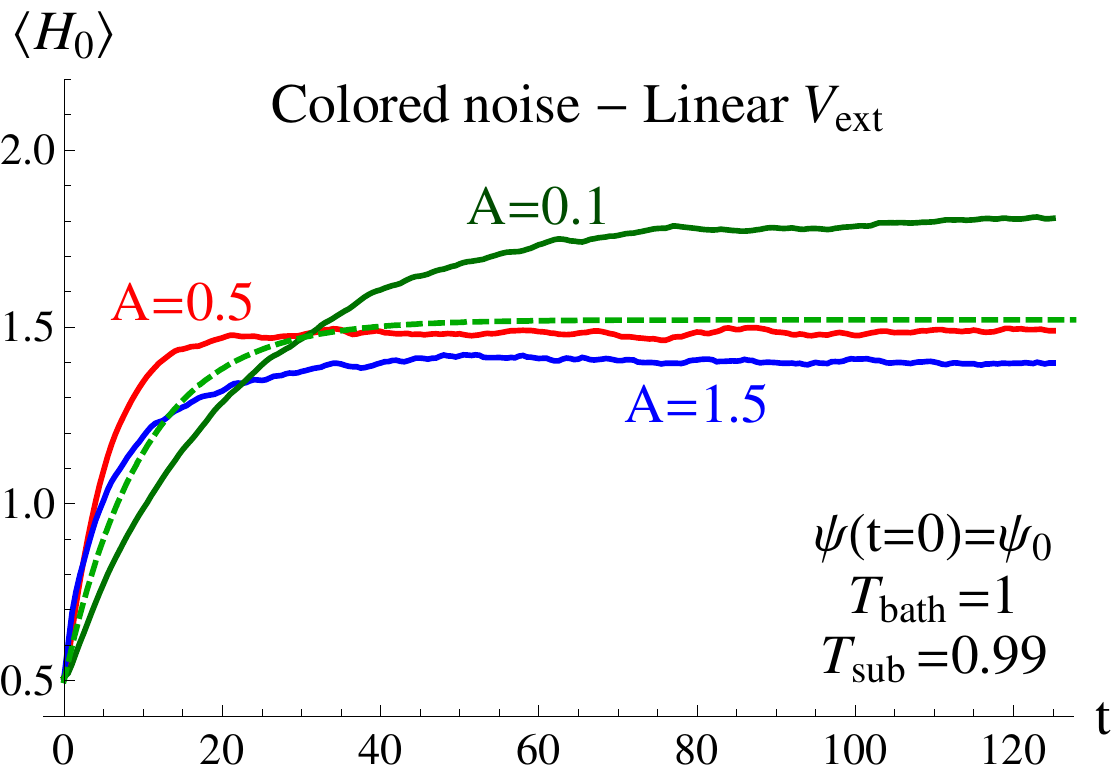}
 \else
 \includegraphics[height=50mm]{AverageEnergyEvolLinColor.pdf}
 \fi
}
   \caption{\label{fig: AvEnLinCol}
   \small   Numerical average energy $\langle H_0\rangle$ evolutions for different friction coefficients $A$ (solid curves) and the theoretical evolution given by (\ref{SenEnerEvol}) with $T_{\rm bath}=1$, $\langle H_0\rangle(t\rightarrow\infty)=1.52$, and $A=0.1$ (dashed curve).}
\end{figure} 
\begin{figure}[!h]\label{fig: AsymLinPnCol}
 \centerline{
 \iffigsdirectory
 \includegraphics[height=50mm]{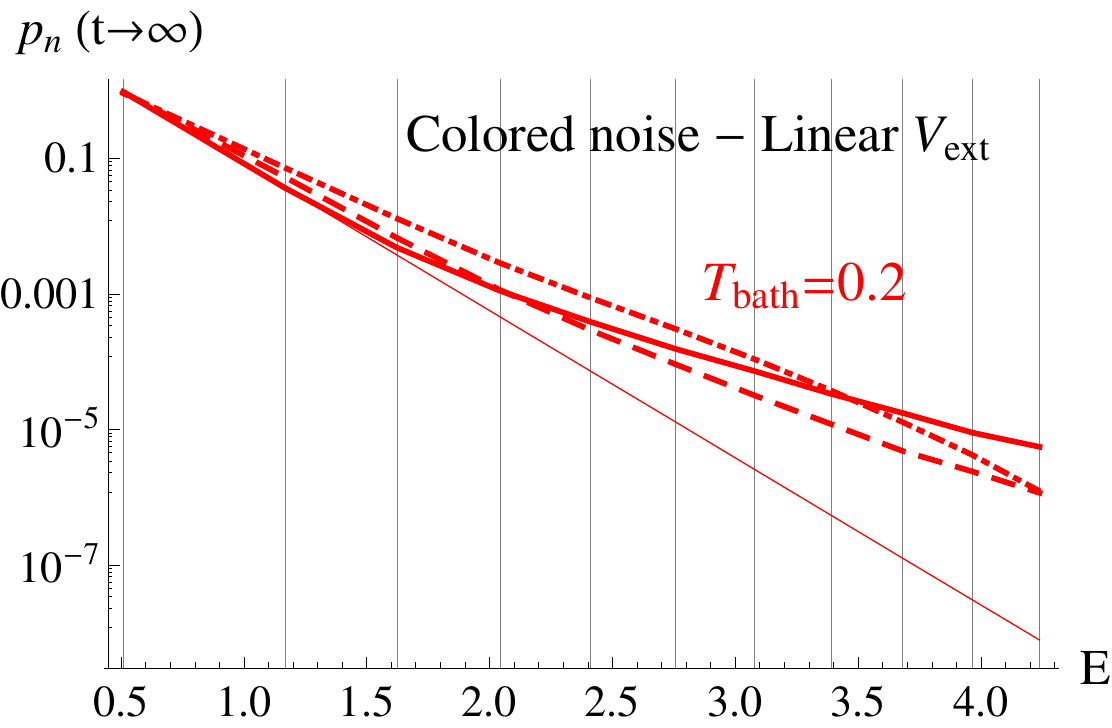}
 \includegraphics[height=50mm]{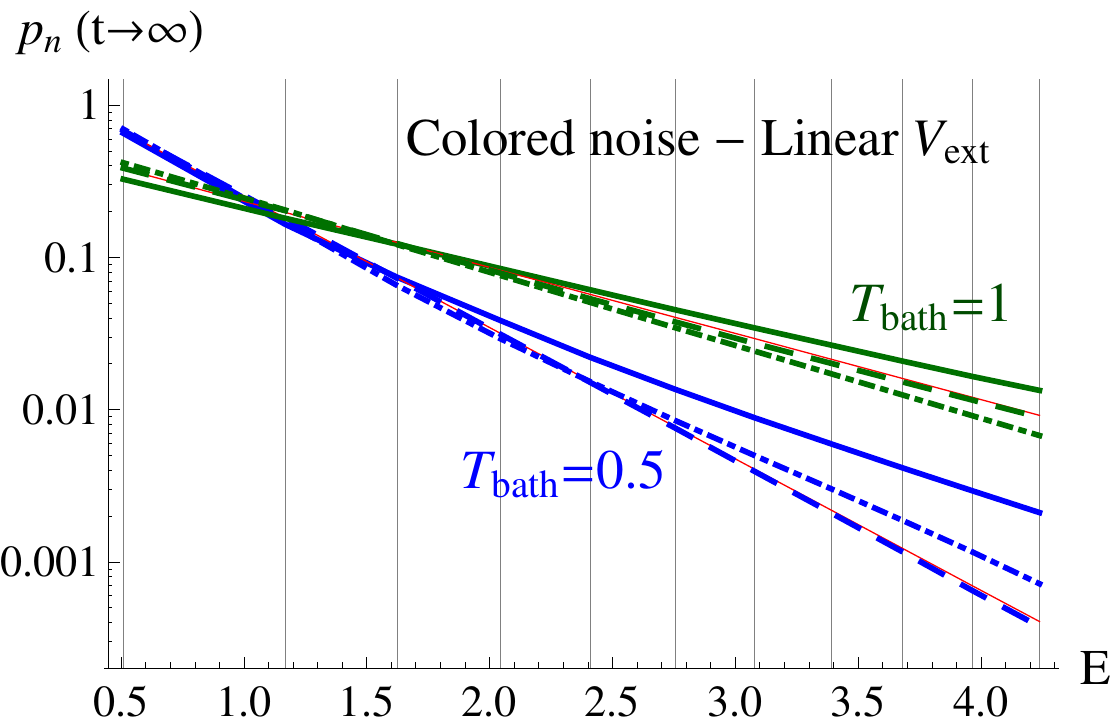}
 \else
 \includegraphics[height=50mm]{LinWeightsAsympT02Col.pdf}
 \includegraphics[height=50mm]{LinWeightsAsympT05and1Col.pdf}
 \fi
}
   \caption{\label{fig: AsymLinPnCol}
   \small   The asymptotic distributions of the eigenstate weights $p_{n=0,...10}$ (joined by lines) function of the eigenenergies $E_{n=0,...10}$ (vertical lines), obtained with different friction coefficients $A=0.1$ (solid lines), $A=0.5$ (dashed lines) and $A=1.5$ (dot-dashed lines) and temperatures $T_{\rm bath}=0.2$ ({\it left}), $T_{\rm bath}=0.5$ and $1$ ({\it right}). They are compared to the corresponding ``ideal" Boltzmann distributions $\propto e^{-E/T_{\rm bath}}$ (thin lines).}
\end{figure} 

As shown in Fig.~\ref{fig: AvEnLinCol}, the $\langle H_0 \rangle$ average energies show similar features to the ones obtained with the white noise (Fig.~\ref{fig: AvEnLinW}), although one
observes some overshooting of the expected value (1.52) at
small coupling. The asymptotic distributions of the weights are found to be independent of the initial state and are displayed in Fig.~\ref{fig: AsymLinPnCol}. They are given with the same values of $\{A,T_{\rm bath}\}$ as for the harmonic oscillator case (Fig.~\ref{fig:AsymHarmoPnCol}) which constitutes the reference
point for our analysis. In that case, one has found two overlapping regimes for which the Boltzmann distribution is recovered, namely the high temperature regime $T_{\rm bath}\gg 1$ and the weak coupling regime $A\ll \{1,T_{\rm bath}\}$. For the linear potential, one observes as well a good agreement with the Boltzmann distribution in the high temperature regime, as illustrated by the $T_{\rm bath}=1$ set of curves in the right panel of Fig.~\ref{fig: AsymLinPnCol}, with larger dispersion than for the harmonic potential, though. 

For a fixed $T_{\rm bath}\lesssim 1$, large deviations are always found in the strong coupling regime $A\gtrsim 1$, especially when $T_{\rm bath}\ll 1$ (as illustrated by the dashed curve on the left panel). In the weak coupling regime,
however, the Boltzmann distributions are only matched for a {\em limited number of low lying eigenstates} which appears 
to be reduced as compared to the harmonic case.
It is still an open question whether one has $\lim_{A\rightarrow 0} p_n \propto e^{-E_n/T_{\rm bath}}$ in the weak coupling limit (a convergence that could be at best non uniform) or if finite deviations from the Boltzmann distribution survive. To conclude our comparison, one should note that the 
distributions of weights for the linear potential show overall richer pattern at small $T_{\rm bath}$, as for instance non monotonous dependences vs the friction coefficient or similar alternating patterns as in the white noise case (see Fig.~\ref{fig: AsymLinPn}) with however lighter oscillations.

 Despite these discrepancies, the relation $T_{\rm sub}$ vs. $T_{\rm bath}$ (Fig.~\ref{fig: ColorLinTsubTbath}), obtained by focusing on the lowest excited states, is interestingly close to the one obtained with the harmonic potential. In the weak coupling regime, one recovers in particular $T_{\rm sub}\approx T_{\rm bath}$, except for $T_{\rm bath}\gtrsim 2$ where one naturally recovers the white noise results, with $T_{\rm sub}>T_{\rm bath}$, as already observed on Fig.~\ref{fig: TeffvsT}) . These observations confirm the rather general nature of the colored noise (\ref{QNFordNormProd}), which might thus be combined with a wider class of potentials and used in a good approximation for thermalization studies especially in the weak coupling case.

 \begin{figure}[!h]\label{fig: ColorLinTsubTbath}
  \centerline{
  \iffigsdirectory
  \includegraphics[height=50mm]{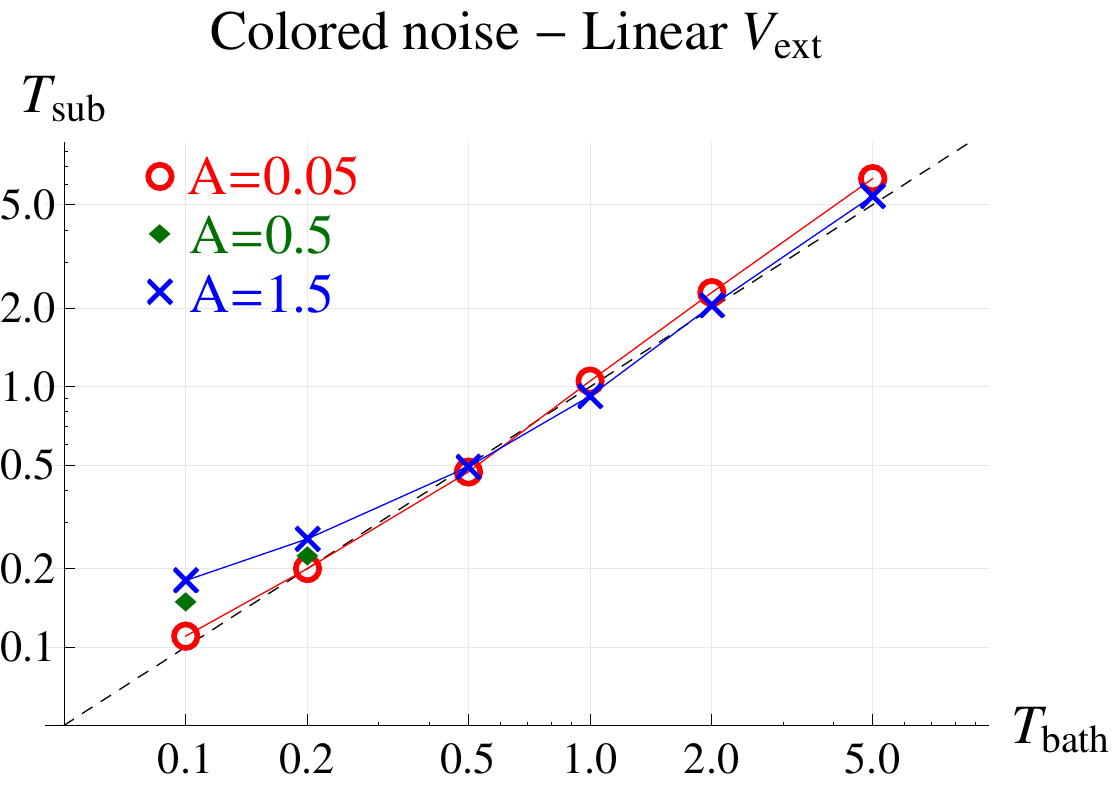}
  \else
  \includegraphics[height=50mm]{ColoredLinTeffvsT.pdf}
  \fi
 }
    \caption{\label{fig: ColorLinTsubTbath}
    \small   Asymptotic subsystem temperature $T_{\rm sub}$ as a function of the bath temperature $T_{\rm bath}$ for three different friction coefficients: $A=0.05$ (red circles), $A=0.5$ (green diamonds) and $A=1.5$ (blue crosses) corresponding respectively to a weak, intermediate and strong coupling. The dashed line corresponds to the ideal case $T_{\rm sub}=T_{\rm bath}$.}
 \end{figure} 
 %

\subsection{Application to an interesting subatomic system}\label{Section5LinearCol_appli}
Although one has strong evidence that protons, neutrons and other hadrons are made of more fundamental objects -- the quarks and the gluons -- these constituents are usually confined inside those hadrons. At high temperature or density, the coupling constant is reduced and the quarks and gluons are expected to be found in a deconfined phase (the so called quark-gluon plasma (QGP)), probably achieved for a short lapse of time in the early universe. Nowadays, it is thought that such a state could be formed and investigated at best in ultra relativistic heavy ion collisions (URHIC) taking place in large ion colliders, as for instance RHIC and LHC located respectively at Brookhaven National Laboratory and at CERN. One of the possible signatures of the QGP is the melting of all hadrons made of $c \bar{c}$ and $b \bar{b}$ quark-antiquark pairs -- the so-called {\it quarkonia} otherwise quite robust in less extreme conditions -- resulting in a suppression of their production as compared to a situation were no QGP would have been formed. Such a mechanism was postulated in 1986 by Matsui and Satz~\cite{Matsui:1986} and was indeed observed later on, but the analysis is however not quite conclusive (see~\cite{SaporeGravis:2015} for a recent review), one of the main reason, on the theory side, being that the dynamics of those quantum object is not properly taken into account in most models. It is the purpose of this section to investigate whether the SLE could be appropriate to deal with such a system. 

Quarkonia can be described in non-relativistic quantum mechanics through phenomenological potentials, the most simple of them being the Cornell potential (see~\cite{Eichten:1995} and references therein):
\begin{equation} 
V_{\rm Cornell}(r):= - \frac{\kappa}{r}+ s r\,,
\label{eq:cornell_potential}
\end{equation}
where $r$ is the distance between the quark and the antiquark,
$\kappa=0.52\hbar c$ and $s \approx$ 1 GeV/fm is the string tension. It is thus the relative coordinates which should be taken as the fundamental degrees of freedom of the SLE. To make the link with the simple 1D linear potential studied in sections
\ref{Section4LinearWhite} and \ref{Section4LinearCol}, one can 
note that the Coulomb part $\propto - \frac{\kappa}{r}$ in 
(\ref{eq:cornell_potential}) mainly acts on the ground state but has much smaller influence on excited states, especially for $c\bar{c}$ bound states (charmonia). For the sake of the analysis, we will thus associate $V_{\rm Cornell}(r)$ to a linear 1D potential with $K_l = 2s \approx  2~{\rm GeV/fm}$. As for the quark mass, the optimal parameters extracted from spectroscopy analysis~\cite{Eichten:1995} are $m_c\approx 1.84~{\rm GeV}/c^2$
and $m_b\approx 5.2~{\rm GeV}/c^2$. Concentrating on the charmonia 
case, one gets a reduced mass $\mu\approx 
0.9~{\rm GeV}/c^2$ leading to  
\begin{equation}
\hbar \omega_0 =
 \sqrt[3]{\frac{(\hbar c K_l)^2}{\mu c^2}}\approx 
  \sqrt[3]{\frac{(0.2\times 2)^2}{0.9}}\approx 0.55~{\rm GeV}\,,
\end{equation}
as $\hbar c\approx 0.2~{\rm GeV\cdot fm}$. Typical temperatures reached in most energetic URHIC are of the order of 
$k T_{\rm bath} \in [0.15~{\rm GeV}$, $0.6~{\rm GeV}]$, resulting in 
$\frac{k T_{\rm bath}}{\hbar \omega_0}\in [0.3,1.1]$.

As for the friction coefficient of heavy quarks interacting 
with a quark gluon plasma, one can find several estimations in the literature, and we will here refer to the calculation of one of the authors~\cite{Gossiaux:1998}. In this work, it was found that $A\approx 1.5 kT_{\rm bath} ({\rm c/fm})$, with $k T_{\rm bath}$ is expressed in GeV and where the increase with the temperature results from the increasing density of interacting particles (quarks and gluons) which are the ground for friction. Consequently, one finds 
$\hbar A\approx 1.5 kT_{\rm bath} \frac{\hbar c}{\rm fm}\approx 
0.3 k T_{\rm bath}$, implying that one never encounters a breakdown of the Brownian hierarchy, even at the smallest temperatures achieved in URHIC. As $\hbar A \lesssim k T_{\rm bath}\lesssim  1.5\,\hbar\omega_0$, we infer that this system lies in the intermediate coupling regime. Examination of Fig.~\ref{fig: AsymLinPnCol} reveals that states up to the 3rd excited one ($E_3
\approx 2 \hbar \omega_0)$ should be correctly described by the SLE, which represents most of the charmonia spectroscopy. All together, we conclude that the SLE is a relevant approach to study the dynamics of quarkonia formation in URHIC and probe more accurately the QGP formed in these collisions. 

\section{Discussion and conclusion}\label{Conclu}

For the purpose of finding an effective formalism suitable to phenomenological applications of open quantum systems, we have focused on the Schr\"odinger--Langevin equation (\ref{SLeq}). Its nonlinear friction term is commonly believed to maintain the stationarity of the excited states of the uncoupled Hamiltonian $H_0$. We have shown in Sec.~\ref{PolarPres} that the Madelung/polar transformation of the wave function leads to a nonzero damping for these states. In this way, we have reconciled the SLE with the intuitive expectation that the dissipation process should act on any state in order to bring the subsystem to its ground state.

We have then focused on the question of the thermal relaxation dynamics given by the SLE for two different potentials and with two different noise operators, taken as c-numbers: the white noise (\ref{QNRK}) -- which has been derived by Senitzky \cite{Senitzky} and subtracted by its term of ground state fluctuations -- and the colored noise (\ref{QNFordNormProd}) derived by Ford, Kac and Mazur \cite{Ford:1965}. 

We first considered the case of a harmonic potential for which most of the results regarding asymptotic states and distributions could be established analytically for both kind of noises. When the subsystem undergoes a white noise, the SLE has demonstrated its ability to bring any initial state to the thermal equilibrium of statistical mechanics (i.e.~Boltzmann distributions of the uncoupled subsystem energy states), irrespective of the coupling strength and of the bath temperature -- confirming the assumption made by Messer \cite{Messer:1979} -- although such equilibration is generally expected only in the weak coupling limit (as explained in the introduction). For a colored noise, exact thermalization toward the Boltzmann distribution was established only in the large temperature regime ($k T_{\rm bath}\gg \hbar \omega_0$) or in the so-called weak coupling regime ($A\ll \{\omega_0,k T_{\rm bath}/\hbar\}$) where the Brownian hierarchy is satisfied.
Outside of these regimes, it has been shown that most of the deviations with respect to the Boltzmann distribution $\exp(-E_n/k T_{\rm bath})$ could be accommodated by introducing some effective temperature $T_{\rm sub}$ corresponding to the subsystem internal equilibration. The disagreements between $T_{\rm sub}$ and the bath temperature $T_{\rm bath}$ (input of the noise) observed at low temperatures or for large friction coefficient $A$ can be attributed to the breaking of the Brownian hierarchy. 

The study for the case of a subsystem submitted to a linear potential
was performed thanks to numerical investigations. While a rather clean equilibration was found in the high temperature regime, genuine non-Boltzmannian behaviors for higher excited states (which cannot be accommodated by a simple change of the temperature $T_{\rm bath}\rightarrow T_{\rm sub}$) and stronger friction coefficient dependences have been observed at low and medium temperatures for both kind of noises. Concentrating on a smaller subset of low lying eigenstates, the colored noise has nevertheless led to better results in the sense of statistical mechanics, confirming its more universal nature. In particular, the effective temperature $T_{\rm sub}$ was found to follow similar behavior as the one introduced in the harmonic potential case and to converge toward $T_{\rm bath}$ in the weak coupling regime. We thus conclude that the colored noise should be used preferentially, especially at low temperatures (under the condition that the Brownian hierarchy is preserved).

The SLE and the quasiclassical Langevin equation seem therefore to share a common difficulty in the description of dissipative evolutions outside the nearly harmonic and free potential cases \cite{Weiss:2012} and especially outside the classical high temperature limit \cite{Hanggi:2005}. Nevertheless, further analyses -- such as in quantum tunneling \cite{Eckern:1990} -- would be required to establish a common behavior. 

If one focuses on applications where only the lower states are considered, a phenomenological model could be to rectify the observed differences between $T_{\rm sub}$ and $T_{\rm bath}$ by choosing an effective heat-bath temperature $\tilde{T}_{\rm bath}$ such as to reach the desired subsystem temperature $T_{\rm sub}=T_{\rm bath}$. It just requires the proper knowledge of the $T_{\rm sub}$($T_{\rm bath}$) function as displayed in Figs.~\ref{fig: HColTeffvsT},~\ref{fig: TeffvsT} and \ref{fig: ColorLinTsubTbath}. Though one has to adapt this rescaling to each situation, the SLE can thus be considered, in a good approximation, as a possible effective alternative to quantum master equations and SSE for phenomenology of complex quantum systems (such as ions quantum transport, thermalization of quarkonia in a quark-gluon plasma,
of nucleons in a nucleus,\ldots). Dealing with the full hierarchy of states in the general case would possibly require either the use of the colored noise correlation (\ref{QNFord}), of a q-number noise operator or of a more refined quantum treatment of the subsystem interactions with the heat bath. Establishing an Einstein relation specific to the SLE would also be an interesting perspective for this field. 

It should be noted that our analysis relies on the hypothesis that the asymptotic distribution of subsystem-eigenstates weights $p_n$ must be Boltzmannian whatever the potential and the coupling strength to the rest of the system (the heat bath). To our knowledge, such an assumption has not been universally established from fundamental principles (i.e.~starting from the distribution of the full-system eigenstates and tracing out the heat-bath degrees of freedom) and should be considered more thoroughly in future studies as it could partially alter our conclusions. 

\appendix
\section{Numerical generation of noises}
\label{appendixNoise}\label{QNnum}
We describe here a numerical method used to sample numerically colored noises characterized by finite autocorrelation times such as (\ref{QNFordNormProd}). In a first step, one generates stochastic stationary Gaussian random variables $\hat{r}_j$ with zero average and correlation $\langle \hat{r}_j\,\hat{r}_{j'}\rangle=\Delta t\,\delta_{jj'}$, where $\Delta t$ is the time step of the numerical scheme. Then, one builds the Gaussian random force $\hat{F}$ at a time $t_i$ -- and assumed to be constant over the time step $[t_i,t_i+\Delta t]$ -- from the weighted sum
\begin{eqnarray}\label{FandW}
\hat{F}_i=\sum_{j=-\infty}^{+\infty}W_{i-j}\,\hat{r}_j\,,
\end{eqnarray}
where the weights $W_{i-j}$ depend only on the difference $i-j$ to guarantee the stationarity of the process. Then, the average of $\hat{F_i}$ is null and its covariance is given by
\begin{eqnarray}\label{Fcov}
\left<\hat{F}_i\,\hat{F}_{i'}\right>=\sum_{j,j'=-\infty}^{+\infty}W_{i-j}W_{i'-j'}\langle \hat{r}_j\,\hat{r}_{j'}\rangle
=\sum_{j=-\infty}^{+\infty}W_{i-j}W_{i'-j}\Delta t\,,
\end{eqnarray}
which, in the continuous limit $\Delta t \rightarrow 0$, becomes 
\begin{eqnarray}\label{covWtau}
\left<F_R(t)F_R(t')\right>=\int_{-\infty}^{+\infty}\mathcal{W}(t-t'')\,\mathcal{W}(t'-t'')\,dt''\,,
\end{eqnarray}
with $W_{i}=\mathcal{W}(t_i)$ for a given time step $\Delta t$.
Then, one easily shows that the Fourier transform of $\mathcal{W}$ is just the square root of the power spectrum $P(\omega)$ of the retained noises, i.e.~ 
\begin{eqnarray}
P(\omega)&=&|\tilde{\mathcal{W}}(\omega)|^2
\end{eqnarray}
with
\begin{eqnarray}\label{PSc}
P(\omega)&=&2mA\,\frac{\hbar \omega}{\exp(\hbar \omega/kT_{\rm bath})-1}\,, 
\end{eqnarray}
for the colored quantum noise (\ref{QNFordNormProd}) and 
\begin{eqnarray}\label{PSw}
P(\omega)&=&\lim_{\sigma\rightarrow 0}\, \,B \,\exp\bigg(-\frac{1}{2}\sigma^2\omega^2\,\bigg)\,, 
\end{eqnarray}
for a Gaussian $C$ with autocorrelation time $\sigma$. the white quantum noise (\ref{QNRK}) is obtained from the latter when $\sigma\rightarrow 0$, but in practice it is sufficient to take $\sigma$ much smaller then the typical times governing the subsystem evolution. Then, one gets explicitly
\begin{eqnarray}\label{Wtau}
\mathcal{W}(\tau)=\frac{1}{\pi}\int_{0}^{\infty} \sqrt{P(\omega)} \cos(\omega\tau)d\omega\,
\end{eqnarray}
and then the stochastic variables $\{\hat{F}_i\}$ through equation (\ref{FandW}). 

\begin{figure}[H]\label{fig: ColoredNoiseEx}
 \centerline{
 \iffigsdirectory
 \includegraphics[height=50mm]{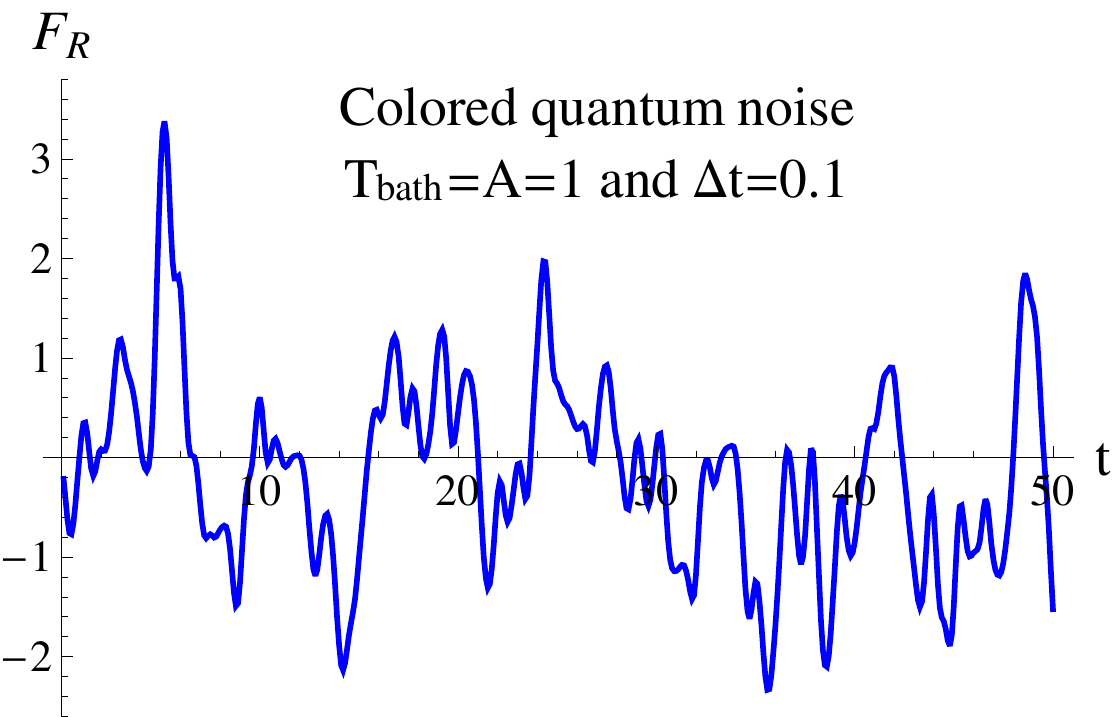}
 \includegraphics[height=50mm]{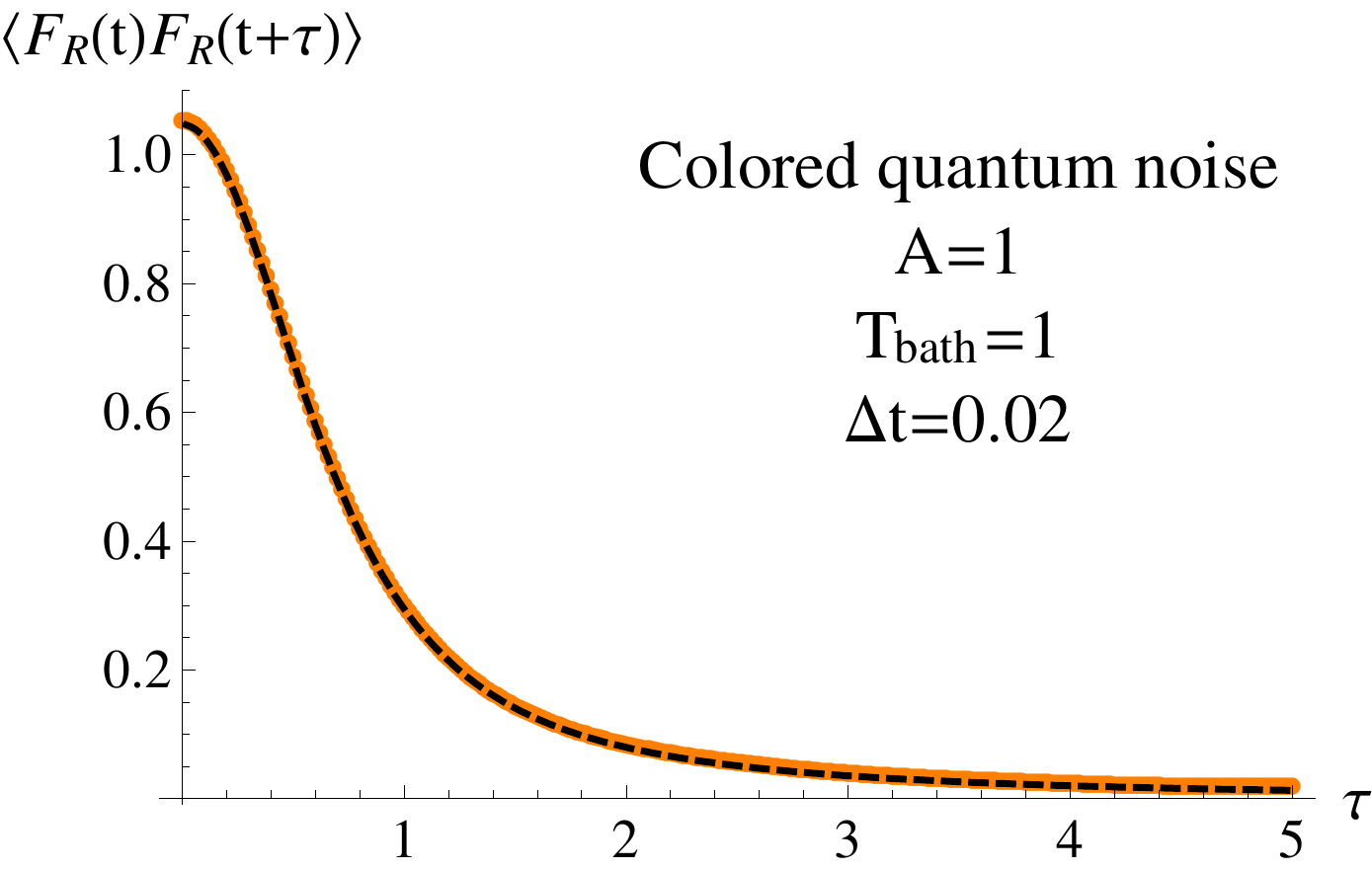}
 \else
 \includegraphics[height=50mm]{ColoredNoiseReal.pdf}
 \includegraphics[height=50mm]{ColoredNoiseCovariance.pdf}
 \fi
}
   \caption{\label{fig: ColoredNoiseEx}
   \small   {\it Left}: Example of one colored noise (\ref{QNFordNormProd}) realization obtained with the described numerical method. {\it Right}: Corresponding analytical (dashed black curve) vs. numerical (orange dots) covariances over time.}
\end{figure} 

In Fig.~\ref{fig: ColoredNoiseEx} (left), an example of a colored noise (\ref{QNFordNormProd}) realization obtained with the described numerical method is shown. In Fig.~\ref{fig: ColoredNoiseEx} (right), the corresponding numerical correlation over time is successfully compared to the analytical expectation.  

Besides, one can easily show that the variables defined in this way are Gaussian. Similar algorithms can be found in the literature and have been successfully used in SSE and other formalisms (see \cite{Biele:2014} and references therein).

\section{Evolution of a Gaussian wave packet under the SLE with a harmonic potential}
\label{appendixGaussianevolve}

\begin{figure}[H]
\begin{center}
\iffigsdirectory
\includegraphics[width=6cm]{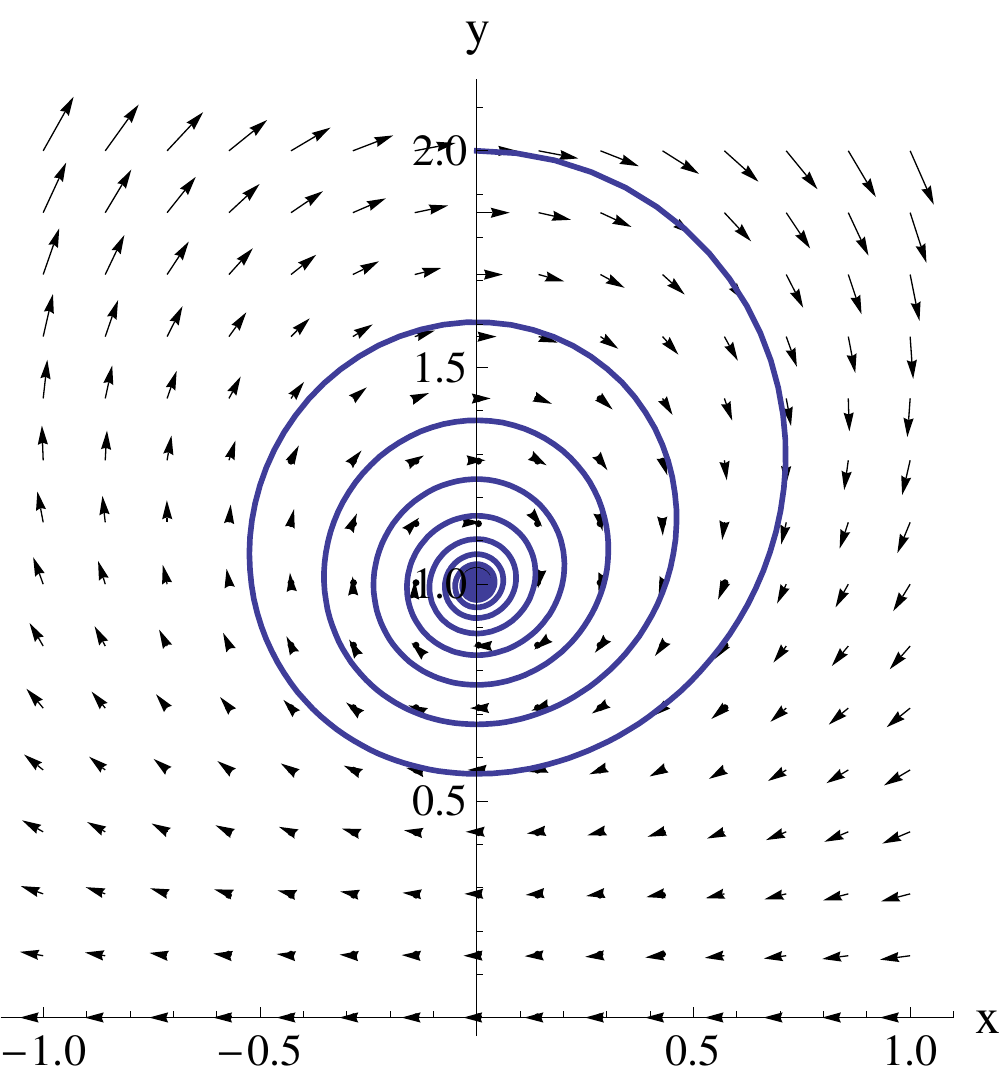}
\else
\includegraphics[width=6cm]{vector_field.pdf}
\fi
\end{center}
\caption{Illustration of the vector field $\vec{f}$ defined in equation (\ref{eq:evol_gamma_harmonic_red_2D}) for $\tilde{A}=1/4$; the black disk 
corresponds to the stationary point $z_{\rm st}=(0,1)$. The spiral curve represents a solution of the equations of motion
(\ref{eq:evol_gamma_harmonic_red_2D}) for initial conditions
$(x(0)=0,y(0)=2)$.}
\label{fig:vector_field}
\end{figure}

The evolution of a Gaussian wave packet under the SLE has been extensively discussed by several 
authors (see \cite{Haas:2013} and references therein). Here we derive anew the essential results for our study,
concentrating on the asymptotic shape of the solution. We consider the 1D SLE with a harmonic potential:
\begin{equation}
i\hbar \frac{\partial \psi}{\partial t}=
-\frac{\hbar^2}{2 m}  \frac{\partial^2\psi}{\partial x^2}
+ \hbar A \left(S(x) -\langle S\rangle \right) \psi - 
F_R(t) x\,\psi  + \frac{K}{2} x^2\, \psi \,,
\end{equation}
where $K=m \omega^2_0$, $A$ is expressed in ${\rm s}^{-1}$ and plays the role of a relaxation rate, while $S$ is the phase of $\psi$ and $F_R(t)$ is a stochastic force. We consider the following {\em Ansatz} as a specific class
of solutions
\begin{equation}
\psi_{A} \propto \exp\left[\frac{i}{\hbar} \left(\alpha(t) (x-x_{\rm cl}(t))^2 +
p_{\rm cl}(t) (x-x_{\rm cl}(t)) +\gamma(t)\right)\right]\,,
\label{eq:gaussian_ansatz_psi}
\end{equation}
where $\alpha$ governs the wave packet width (with $\Im(\alpha)>0$), $x_{\rm cl}$ is the centroid in usual space and 
$p_{\rm cl}$ is the centroid in momentum space, both assumed to be real.
$\gamma$ combines a phase factor (real part) and an normalizing factor (imaginary part). The l.h.s. of the SLE gives
\begin{eqnarray}
i\hbar \frac{\partial \psi_A}{\partial t}&=&
-\left[\dot\alpha(t) (x-x_{\rm cl}(t))^2 +
2 \alpha(t) \dot x_{\rm cl}(t)  (x-x_{\rm cl}(t))+
\dot p_{\rm cl}(t)  (x-x_{\rm cl}(t))-
\right.
\nonumber\\
&& \left.p_{\rm cl}(t) \dot x_{\rm cl}(t) +\dot \gamma(t)\right] \psi_A\,.
\end{eqnarray} 
As for the r.h.s, one has 
\begin{equation}
-\frac{\hbar^2}{2 m}  \frac{\partial^2\psi_A}{\partial x^2} = 
\frac{1}{2m} \left\{-2 i \hbar\alpha(t)  + \left[2 \alpha(t)  (x-x_{\rm cl}(t)) + 
p_{\rm cl}(t) \right]^2 \right\}  \psi_A
\end{equation}
while the friction term leads to 
\begin{equation}
\hbar A \left(S(x) -\langle S\rangle \right) \psi_A=
A\left[\Re(\alpha(t)) (x-x_{\rm cl}(t))^2
+ p_{\rm cl}(t) (x-x_{\rm cl}(t))\right] \psi_A\,.
\end{equation}
Both l.h.s. and r.h.s. correspond to $\psi_A$ multiplied by second degree
polynomials. Equating terms $\propto x^2$ leads to a first equation:
\begin{equation}
\boxed{\dot{\alpha}=-\frac{2 \alpha^2}{m} - A \Re(\alpha)
-\frac{m \omega_0^2}{2}}\,.
\label{eq_alpha_harmcase_case}
\end{equation}
Writing 
\begin{equation}
\frac{m\omega_0^2}{2} x^2 = \frac{m\omega_0^2}{2}
\left[(x-x_{\rm cl}(t))^2+2 x_{\rm cl}(t) (x-x_{\rm cl}(t))+
x_{\rm cl}^2(t) \right],
\end{equation}
all terms $\propto (x-x_{\rm cl}(t))^2$ cancel provided equation
(\ref{eq_alpha_harmcase_case}) is satisfied, and we thus equate
 terms $\propto (x-x_{\rm cl}(t))$:
\begin{eqnarray}
-\left[2\alpha(t) \dot x_{\rm cl}(t) +
\dot  p_{\rm cl}(t) \right] (x-x_{\rm cl}(t))&=&\\
&&\hspace{-4cm}\left[\left(\frac{2\alpha(t)}{m} + A \right) p_{\rm cl}(t)
- F_R(t) + m\omega_0^2 x_{\rm cl}(t)\right](x-x_{\rm cl}(t)) \,,\nonumber
\end{eqnarray}
where we have used $F_R(t) x =F_R(t)(x-x_{\rm cl}(t)) +
F_R(t) x_{\rm cl}(t)$. As $\alpha(t)\in 
\mathbb{C}$ while other quantities are real, real solutions for $x_{\rm cl}$ and $p_{\rm cl}$
are obtained by identifying the powers of $\alpha$ in both 
members of the equation, leading to
\begin{equation}
\boxed{\dot x_{\rm cl}=\frac{p_{\rm cl}}{m}
\quad\text{and}\quad
\dot p_{\rm cl}= - Ap_{\rm cl}+ F_R(t)-  m\omega_0^2 x_{\rm cl} } \,,
\label{eq:append_classical_motion}
\end{equation}
which are just the equations of motion for a classical particle evolving
in some harmonic potential under
the action of a friction term and a fluctuating force. Finally, 
we are left with the constant terms in the polynomial: 
\begin{equation}
p_{\rm cl}(t)\dot x_{\rm cl}(t) -\dot\gamma(t)=
-\frac{i \hbar  \alpha(t)}{m}
+ \frac{p^2_{\rm cl}(t)}{2m} - F_R(t) x_{\rm cl}(t) 
+\frac{m\omega_0^2}{2} x_{\rm cl}^2(t) \,,
\end{equation}
that is, using equation (\ref{eq:append_classical_motion})
\begin{equation}
\dot{\gamma}(t)=
\frac{i \hbar  \alpha(t)}{m}
+ \frac{p^2_{\rm cl}(t)}{2m} + F_R(t) x_{\rm cl}(t) 
-\frac{m\omega_0^2}{2} x_{\rm cl}^2(t) \,.
\end{equation}
Expressing $\alpha=\Re(\alpha) + i \Im(\alpha)$ as well as 
$\gamma=\Re(\gamma) + i \Im(\gamma)$ and equating real 
and imaginary quantities leads to 
\begin{equation}
 \frac{d\Re(\gamma)}{dt}=-\frac{\hbar \Im (\alpha)}{m}+
   \frac{p^2_{\rm cl}(t)}{2m} + F_R(t) x_{\rm cl}(t) 
   -\frac{m\omega_0^2}{2} x_{\rm cl}^2(t)
\end{equation}
and
\begin{equation}
   \frac{d\Im(\gamma)}{dt}=\frac{\hbar \Re(\alpha)}{m} \,.
   \label{eq:evol_gamma_harmonic}
\end{equation}
The latest equation represents nothing but the norm conservation.
Indeed, one has 
\begin{equation}
\int d x |\psi_A|^2
=\int dx\, e^{-\frac{2}{\hbar} \left(\Im(\alpha)(x-x_{\rm cl})^2 + \Im (\gamma) \right)}
\propto \frac{e^{-\frac{2 \Im(\gamma)}{\hbar}}}{\left(\Im (\alpha)\right)^{\frac{1}{2}}}
\end{equation}
and norm conservation reads
\begin{equation}
-\frac{2}{\hbar}  \frac{d\Im(\gamma)}{dt} -\frac{1}{2 \Im(\alpha)}
 \frac{d\Im(\alpha)}{dt}=0\,,
\end{equation}
which can be recovered by combining the imaginary part of
eq.~(\ref{eq_alpha_harmcase_case}), i.e.
\begin{equation}
\Im(\dot \alpha) =-\frac{4\Im(\alpha)\,\Re(\alpha)}{m}\,
\label{eq_alpha_harmcase_case_imag}
\end{equation}
and equation (\ref{eq:evol_gamma_harmonic}). We now turn to the study of
equation (\ref{eq_alpha_harmcase_case}), in particular its asymptotic behavior. It is obvious that this equation has a single stationary point,
namely $\alpha_{\rm st}= i \frac{m\omega_0}{2}$, corresponding to the
ground state width $\sqrt{\frac{\hbar}{m\omega_0}}$. We therefore
express $\alpha$ in units of $\frac{m \omega_0}{2}$ ($\alpha=
z \frac{m \omega_0}{2}$) and $t$ in units of $\omega_0^{-1}$
($t=\tau/\omega_0$) and obtain the reduced equation
\begin{equation}
\frac{dz}{d\tau}=-z^2 - 1 - \tilde{A}
\Re(z)\,,
\label{eq:evol_gamma_harmonic_red}
\end{equation}
where $\tilde{A}=\frac{A}{\omega_0}$. We first discuss the case $A=0$
(no friction), for which the general solution of equation (\ref{eq:evol_gamma_harmonic}) writes.
\begin{equation}
z(\tau)= \frac{z(0)\cos\tau -\sin\tau}
{z(0)\sin\tau -\cos\tau}\,.
\end{equation}
Analysis of this solution shows that $z(t)$ revolves, in the complex plane,  on a circle of center $C= \left(0,\frac{1+z(0)\bar{z}(0)}{2\Im(z(0))}\right)$ and of radius 
\begin{equation}
r(x,y)=\frac{\sqrt{(1+z^2(0))(1+\bar{z}^2(0))}}{2\Im(z(0))}
\end{equation} which is an invariant of the motion. On any of these circles, one has ${\rm max}(\Im(z)) \times {\rm min}(\Im(z))=1$, which implies that the
stationary solution $z_{\rm st}=(0,i)$ is located inside each of them, but {\em does not correspond to the asymptotic behavior of any non-trivial solution}. We now turn to the effect of finite $A$ on the evolution. Using the decomposition $z(t)=x(t) + i y(t)$, the real and
imaginary parts of the reduced 
equation (\ref{eq:evol_gamma_harmonic_red}) write
\begin{equation}
\frac{d}{d\tau}
\left(
\begin{array}{c} 
x \\ y
\end{array}\right)=
\vec{f}(x,y)
\quad\text{with}\quad
\vec{f}(x,y)=
\left(\begin{array}{c}
y^2-x^2 - 1 - \tilde{A} x
\\
-2 x y
\end{array}\right)\,.
\label{eq:evol_gamma_harmonic_red_2D}
\end{equation}
Even if this system admits no analytic solution to our knowledge, an
illustration of $\vec{f}$ in the form of a vector field plot (see Fig.~\ref{fig:vector_field}) reveals its global rotating nature\footnote{Confirmed by ${\rm rot}\vec{f}=-4y <0$ in the upper complex plane.} around the center $z_{\rm st}$.
As compared to the $\tilde{A}=0$ case for which the evolution driven
by $\vec{f}$ takes place on a single circle (preserving the invariant 
$r(x,y)$), one can show that the $-\tilde{A} x$ terms causes some "damping"
of the motion associated with a continuous decrease of $r^2$:
\begin{equation}
\frac{d r^2}{d\tau} =-\frac{\tilde{A} x^2(1+x^2+y^2)}{y^2}\,.
\end{equation}
This prevents $z(t)$ to stay on any given circle. Instead, solutions for finite $\tilde{A}$ are inward spirals which all ends at $z_{\rm st}$, as illustrated by the solid line in 
Fig.~\ref{fig:vector_field} :
\begin{equation}
\lim_{\tau \rightarrow +\infty} z(\tau)=z_{\rm st}\,.
\end{equation}
This proves that all Gaussian wave packets evolving in some harmonic potential under the SLE ultimately acquire the ground-state width. 

To conclude, we wish to make the connection with the results established
using the hydrodynamic form of the SLE 
(see for instance \cite{Haas:2013} for a recent discussion). In this approach a Gaussian {\em Ansatz} is made for the probability density
\begin{equation}
\rho =\frac{1}{\sqrt{2\pi}a(t)} \exp\left(-\frac{(x-x_{\rm cl})}{2 a(t)^2}\right)\,,
\label{eq:gaussian_ansatz_rho}
\end{equation}
where $a$ satisfies the Pinney equation
\begin{equation}
\ddot{a}+A \dot{a} + \omega_{0}^2 a = \frac{\hbar^2}{4 m^2 a^3}\,.
\label{eq:Pinney}
\end{equation}
Comparing the Gaussian {\em Ansatz} (\ref{eq:gaussian_ansatz_rho})
with our {\em Ansatz} for $\psi$, equation (\ref{eq:gaussian_ansatz_psi}) leads to 
$\frac{1}{a^2}=\frac{4\Im(\alpha)}{\hbar}\Rightarrow -2\frac{\dot{a}}{a^3}=\frac{4 \Im(\dot{\alpha})}{\hbar}\Leftrightarrow
\dot{a}=-\frac{2 \Im(\dot{\alpha}) a^3}{\hbar}$.
Using, the equation of motion (\ref{eq_alpha_harmcase_case_imag}) for the imaginary part of $\alpha$, one obtains
\begin{equation} 
\dot{a}=\frac{8 \Im(\alpha) \,\Re(\alpha) a^3}
{\hbar m}= \frac{2\,\Re(\alpha) a}
{m}\,.
\label{eq:connection_hydro_SL}
\end{equation}
In order to obtain a closed differential equation in $a$, we 
differentiate equation (\ref{eq:connection_hydro_SL}) and use 
the equation of motion (\ref{eq_alpha_harmcase_case_imag}) for the real part of $\alpha$:
\begin{equation} 
\ddot{a}=\frac{2}{m}\,\left(\Re(\dot \alpha) a + 
\Re(\alpha) \dot a\right)=
\frac{2}{m}\,\left[\left(
\frac{2}{m}(\Im(\alpha)^2-\Re(\alpha)^2)- A \Re(\alpha) -
\frac{m \omega_0^2}{2}\right) a+ 
\Re(\alpha) \dot a\right]\,.
\end{equation}
We then perform the final substitutions
$\Re(\alpha) \rightarrow \frac{m \dot a}{2 a}$
and $\Im(\alpha) \rightarrow \frac{\hbar}{4 a^2}$ and indeed
recover equation (\ref{eq:Pinney}).

\section*{Acknowledgments}
We are grateful for the support from TOGETHER project R\'egion Pays de la Loire.


\end{document}